\documentclass[aps,prl,twocolumn,groupedaddress]{revtex4-2}

\usepackage{bm}                         
\usepackage{upgreek}                    
\usepackage{amsmath}                    
\usepackage[sfdefault=fav]{isomath}     

\usepackage{graphicx}                   
\usepackage{dcolumn}                    

\usepackage{color}

\newcommand{\vect}{\mathbfit}
\newcommand{\norm}[1]{\boldsymbol{\mathrm{#1}}}
\newcommand{\tens}{\mathsfbfit}
\newcommand{\nd}{\norm{n}}
\newcommand{\ke}{\kappa_{\text{eff}}}
\newcommand{\p}[1]{\partial_{#1}}
\def\kt{k_\text{B}T}
\newcommand{\fa}{f_\text{a}}

\begin{document}


\title{The mechanics of disclination emergence in 3D active nematics}


\author{Yingyou Ma}
\affiliation{Martin Fisher School of Physics, Brandeis University, Waltham, Massachusetts 02453, USA}

\author{Christopher Amey}
\affiliation{Martin Fisher School of Physics, Brandeis University, Waltham, Massachusetts 02453, USA}

\author{Aparna Baskaran}
\affiliation{Martin Fisher School of Physics, Brandeis University, Waltham, Massachusetts 02453, USA}
\email{aparna@brandeis.edu}

\author{Michael F. Hagan}
\affiliation{Martin Fisher School of Physics, Brandeis University, Waltham, Massachusetts 02453, USA}
\email{hagan@brandeis.edu}


\date{\today}

\begin{abstract}
The spontaneous creation of disclinations is a defining characteristic of active nematics, which is rarely observed in equilibrium systems or other active matter systems. Thus, understanding the mechanics of disclinations is crucial for developing reliable continuum theories and practical applications. In this work, we explore this intrinsic mechanics by performing large-scale 3D simulations of a particle-based model of active semiflexible filaments. We investigate the effects of filament stiffness and activity on the collective behavior of active nematics. Analysis of the steady state and the topological properties of initial disclination loops reveals that the system is governed by a single parameter, an activity-dependent effective stiffness. Then, we develop a method to visualize director field orientations in a physically transparent manner during the formation of disclination loops. Based on this, we establish a unified theory for the mechanics of disclination emergence, across the range of bend and twist. This disclination analysis framework can also be applied to diverse other 3D liquid crystal systems. 
\end{abstract}

\maketitle

Active materials have the remarkable ability to harness energy at the molecular level to drive self-propulsion, a central phenomenon in organisms such as bacteria~\cite{Dombrowski2004}, birds~\cite{Vicsek1995}, fish~\cite{pitcher2012behaviour}, cells~\cite{annurev-conmatphys-031218-013516}, and humans~\cite{gu_emergence_2025}. The microscopic driving breaks detailed balance and hence gives rise to phenomena explicitly prohibited in equilibrium mechanics~\cite{Marchetti2013, RevABP, Ramaswamy_2017, Seifert_2012, chate_dry_2020, shankar_topological_2022, chate_simple_2006}. Notably, certain active materials are governed by the rich topological characteristics of an order parameter field, which provides a conceptual framework for the dynamics of flow structures in active fluids. An important example is active nematics, where well controlled experiments enable testing the resulting theoretical predictions~\cite{sanchez_spontaneous_2012, duclos_science_2020, zhou_living_2014, Menon_2007}.

Active nematics feature the spontaneous creation and annihilation of topological defects. In two dimensions (2D), the predominant topological features are simple point defects, whose energy, interactions, and shape, among other characteristics, have been extensively investigated in both passive~\cite{kleman_soft_2003} and active~\cite{giomi_defect_2014} systems. In three dimensions (3D), the topological features are disclination lines, exhibiting significantly more intricate topological properties~\cite{mermin_topological_1979, machon_global_2016, alexander_colloquium_2012, mertelj_observation_2004, pollard_contact_2023, duclos_science_2020}, complex structures~\cite{Binysh_2020, duclos_science_2020, simon_topology_2019, schimming_2022, Digregorio_2024, shendruk_twist-induced_2018, long_geometry_2021, bits_2022, Krajnik_spectral_2020} and temporal evolution~\cite{Kralj_loop_dynamics_2023, Binysh_2020, simon_topology_2019, Schimming_2023, Houston_2021, duclos_science_2020, schimming_2022}. These disclinations can be used to control and manipulate flows in active nematic fluids~\cite{jiang_active_2022}. While the mechanics of disclination emergence is well understood in 2D~\cite{giomi_defect_2014}, disclination formation in 3D has only been analyzed in special or simplified cases~\cite{duclos_science_2020, Binysh_2020, shendruk_twist-induced_2018, Nejad_2022}. A comprehensive understanding of the underlying mechanics remains lacking.

\begin{figure}[b]
\includegraphics[width=\columnwidth]{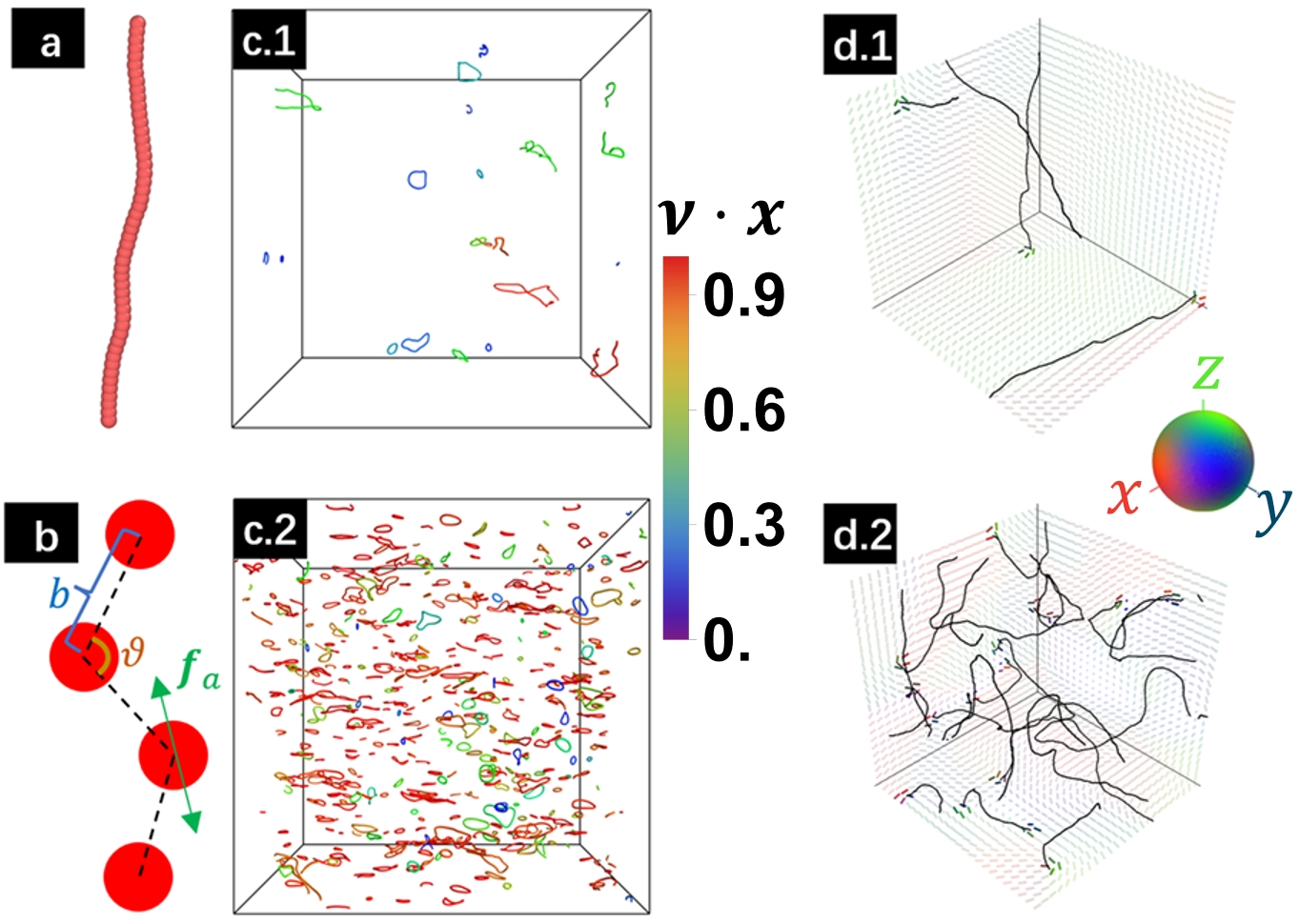}
\caption{\label{fig:pheno} The microscopic model and phenomenology. \textbf{a)} Illustration of an active filament. \textbf{b)} Schematic representation of the active filament model, in which nematic active forces are exerted along the local tangent direction. (see SI Video 1.a-c) \textbf{c)} The initial disclination loops, colored by the alignment between loop's normal $\norm{\upnu}$ and initial order $\norm{x}$.  \textbf{d)} Steady state: zoomed-in views of the network of disclination lines at long times, with directors on the boundary planes color-coded by their orientation. c.1, d.1: activity and stiffness $\fa=1500, \kappa=2$ (effective stiffness $\ke \equiv \kappa/(1+2\fa^2)=167.8$). c.2, d.2: $\fa=1500, \kappa=3.5$ ($\ke=58.8$). Parameter values are reported in dimensionless units defined in the text (see \emph{Model}).
}
\end{figure}

 In this letter, we use large-scale molecular dynamics simulations of active semi-flexible filaments to understand the spontaneous emergence of disclinations from a flat background in 3D active nematics. While this technique has led to fruitful insights in studies of active coarsening systems~\cite{Athani_2024, Winkler_2020, Moore_2020, joshi_interplay_2019, freedman_versatile_2017, palmer_2022},  3D studies are limited to confined systems~\cite{peterson_vesicle_2021, sciortino_active_2025, yan_2025, Henkes_2018} or polar filaments \cite{breoni_giant_2025}. We systematically explore a broad range of filament activity $\fa$ and stiffness $\kappa$ values, both of which control the large-scale system dynamics. Compared to previous studies that used continuum theory~\cite{Digregorio_2024, duclos_science_2020, long_geometry_2021, Kralj_loop_dynamics_2023, Nejad_2022, shendruk_twist-induced_2018}, our particle-based simulations bridge microscopic mechanics, including local alignment and active forces, with the macroscopic continuum field and emergent structures.

Analysis of trajectories shows that both the steady state and the topological properties of initial disclination loops depend  solely on an activity-dependent effective stiffness parameter $\ke$. Furthermore, we develop a framework to understand the complex geometrical structure of the 3D director field and its evolution during the emergence of initial disclination loops. Within this framework, we propose a 2D-analog theory, which extends previous descriptions of the mechanism for pure-bend-induced loops to conditions involving twist. Thus, we establish a unified theory for loop emergence. Finally, by analyzing simulation trajectories we show that loop structures are governed by a competition of timescales for loop nucleation and twist deformations.

\emph{Model.}---To simulate the collective behavior of active semi-flexible filaments in 3D, we extend our previous model for 2D active filaments, which captured the fundamental and essential behaviors and symmetries inherent to active nematics~\cite{joshi_interplay_2019}. We model each filament as a bead-spring polymer, with the monomer coordinates $\vect{r}_i$ governed by Langevin equation:
\begin{equation}
m\ddot{\vect{r}_i}=\vect{f}_\text{a}-\xi\dot{\vect{r}_i}-\nabla_{\vect{r}}U+\vect{R}_i(t)
\end{equation}
where $\vect{f}_\text{a}$ is the active force, $\xi$ is the friction coefficient set to overdamp the dynamics, and $\vect{R}$ is a Gaussian thermal noise with variance $6  \xi \kt$. The potential energy $U$ comprises three components: the FENE bond potential~\cite{fene_1990}, a non-bonded Weeks-Chandler-Anderson volume-exclusion potential~\cite{Weeks1971} which leads to local nematic alignment, and harmonic angle potential $\bar{\kappa}(\pi-\vartheta)^2$ enforcing filament stiffness, where $\vartheta$ is the angle between adjacent bonds and $\bar{\kappa}$ is the elastic modulus. The active force is modeled by a self-propulsion force on each monomer directed along the filament tangent toward its head 
\begin{equation}
    \vect{f}_a=\eta_\alpha(t) \bar{f}_\text{a}\, b \, \norm{t}
\end{equation}
where $b$ is the bond length, $\norm{t}$ is the local unit tangent and $\bar{f}_\text{a}$ is the strength of activity. To keep nematic symmetry, the active force reverses the direction at stochastic time intervals, which effectively switches the head and tail of the filaments. This is reached by $\eta_\alpha(t)$, which changes its value between 1 and -1 at Poisson-distributed time intervals. Fig.~\ref{fig:pheno}(a,b) shows an active filament and a schematic representation of the bead-spring model. To integrate the Langevin dynamics, we modified LAMMPS \cite{LAMMPS} to include the active force \footnote{This modified LAMMPS with the active force is available at https://github.com/mattsep/lammps}. 

In this Letter, we employ the LJ unit system in LAMMPS, where the length, energy, and mass units correspond to the monomer diameter $\sigma$, $\kt$, and monomer mass $m$, respectively. Our simulations used $N\approx2.8\times10^5$ rods in a periodic box with edge length $L=200 \sigma$, bond length $b=0.5 \sigma$, and 50 beads per rod. To probe the instability, we used an aligned nematic as the initial condition. We simulated a wide range of stiffness and activity values (which we report in dimensionless units) $\kappa\equiv\bar{\kappa}/\kt \in [100,1500]$ and $\fa \equiv \bar{f}_\text{a} \sigma/\kt \in [1-10]$,  covering steady states from chaotic to nearly static flow. For analysis, we coarse-grain the system into the tensor order parameter field $\tens{Q}_{ij}=\langle \norm{b}_i\norm{b}_j-\tens{\delta}_{ij}/3\rangle$ where $\norm{b}$ is the normalized displacement between each pair of bonded monomers. The largest eigenvalue and eigenvector of $\tens{Q}$ defines the local order $S$ and director $\nd$. A detailed description of the governing equations and coarse-graining, as well as all parameter values, are presented in the supplementary information (SI).

\emph{Phenomenon.}--- Our system exhibits behavior akin to 3D active nematics experiments composed of microtubles~\cite{duclos_science_2020}. The initially ordered system evolves through the following stages. \textbf{a)} \emph{Instability:} filaments deform and reorient, causing the order parameter $S$ to decrease. \textbf{b)} \emph{Loops:}  the directors undergo significant local distortions, leading to the emergence of disclination loops (Fig.~\ref{fig:pheno}.c), which then grow in size and density, until loops interact with each other. \textbf{(c)} \emph{Isotropy:} the disclinations ultimately evolve to dynamical networks shown in Fig.~\ref{fig:pheno}.d, and the system reaches a steady state that is independent of the initial condition and has stationary statistical properties. SI Videos 1.a-c show a typical simulation trajectory, with each of the stages highlighted. SI Video 2 shows disclination line dynamics in a zoomed region.

\begin{figure}
    \centering
    \includegraphics[width=\columnwidth]{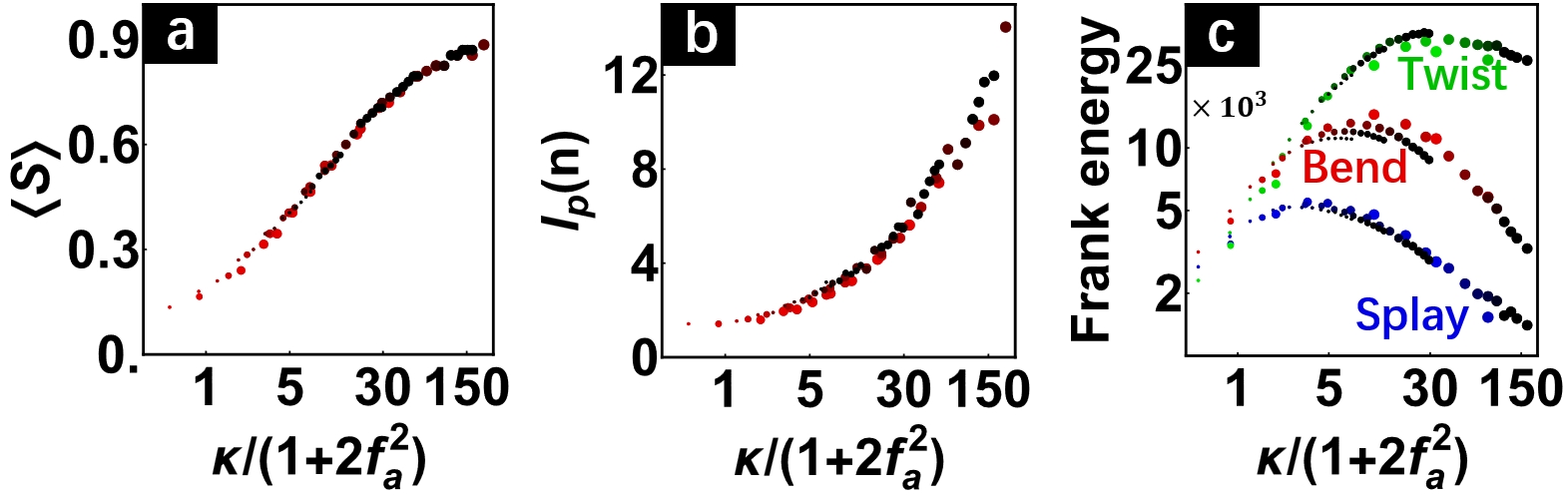}
    \caption{\textbf{(a)} Mean order parameter $S$, \textbf{(b)} auto-correlation length of the director, \textbf{(c)} the total Frank elastic deformation as a function of effective stiffness $\ke$. Darker points indicate higher stiffness and larger points indicate lower activity. }
    \label{fig:collapse}
\end{figure}

\emph{Steady state.}---To characterize the steady states, we present macroscopic quantities commonly used to describe the global properties of nematics, the average order parameter $\langle S \rangle$, the total Frank elastic deformation (i.e., the Frank energy for unit moduli values, see SI section IV), and the auto-correlation length of the director field, as functions of the microscopic parameters $\kappa$ and $\fa$. 
Notably,  Fig.~\ref{fig:collapse} shows that all values collapse onto a single curve parameterized by $\ke \equiv \kappa/(1+2\fa^2)$,  which we denote as the \emph{effective stiffness}. This collapse indicates that $\ke$ predominantly governs the system's statistical properties. Thus, beyond driving disclination formation, activity primarily dissipates in bend  modes, which softens the filaments, consistent with previous observations in 2D~\cite{joshi_interplay_2019}.

We physically interpret the effective stiffness  $\ke=\kappa/(1+2\fa^2)$ as follows. In the zero Reynolds number limit, the elastic stress $\sim\kappa$ balances the combined effects of thermal noise $\sim \kt$ (1 in our units) and active stress. The active stress arises from two factors: the average collision frequency $\sim \fa/\ell$, where $\ell$ is the mean spacing between neighboring monomers, and the collision force strength $\fa$. Consequently, active stress scales as $\fa^2$, effectively reducing the filament bending modulus to $\kappa/(1+ g \fa^2)$, where $g\approx2$ is an empirical coefficient. Details about calculating the statistical properties and the data collapse with $\ke$ are in the SI.

\begin{figure*}
\includegraphics[width=\linewidth]{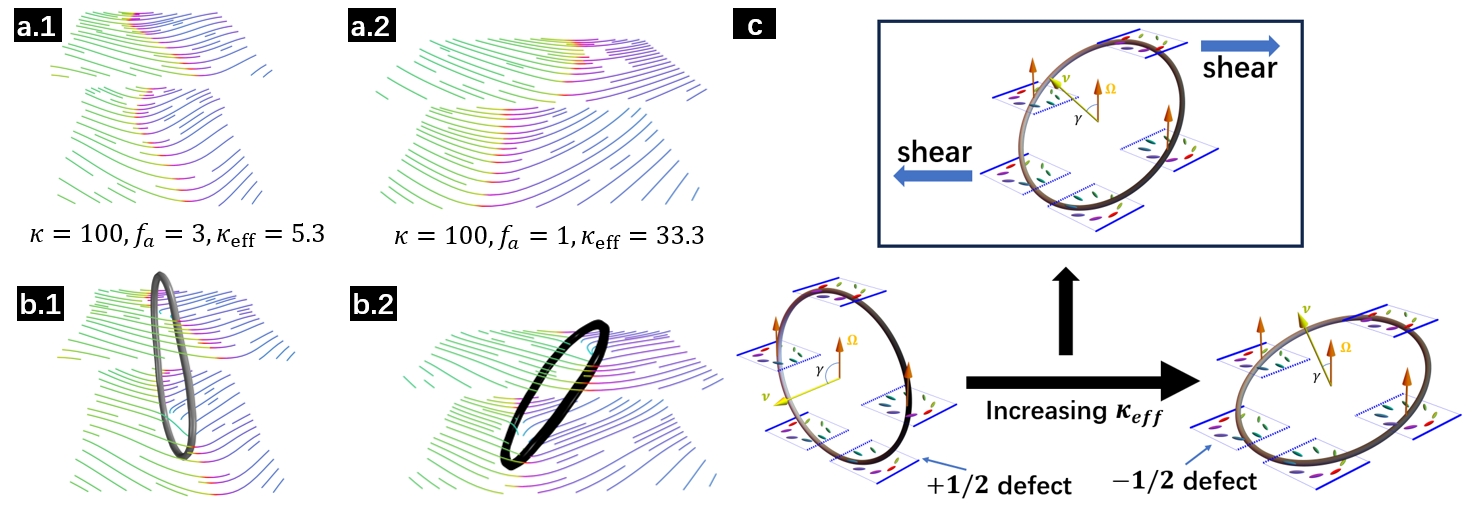}
\caption{\label{fig:observe} Phenomenology of emergence of initial disclination loops. 
\textbf{a)} Visualization of the $\nd$ field projected along the principal planes before the loop's formation from simulations at $\ke$ of 5.3 (left) and 33.3 (right) . \textbf{b)} The initial loops emerge from (a) at a later time $t$. Each principal plane exhibits a pair of bend-induced $\pm1/2$ 2D-like defects. \textbf{(c)} A schematic illustration of the initial loops, to visualize the idealized version of the structures in (b). The diagram highlights how increasing $\ke$ shears the bend structure, inducing additional twist and thereby reducing the angle $\gamma$ between the loop's normal $\norm{\upnu}$ and the rotation vector $\norm{\Omega}$. }
\end{figure*}

\emph{Instability.} ---To investigate the mechanism that leads to the initial disclinations, it is useful to consider a continuum description of the dynamics of our system. As Fig.~\ref{fig:observe}.a shows, consistent with previous findings, ~\cite{Nejad_2022, Elias_2016, sanchez_spontaneous_2012, giomi_defect_2014, Minu_bend_2020}, the bend deformation is the dominant mode. Given that we are simulating a dry system, the dynamics is determined by a balance between Stokesian friction and active stress: $\zeta\vect{v}=\nabla\cdot\tens{\sigma}_\text{a}$, with $\tens{\sigma}_\text{a}=-\alpha\nd\nd$ and $\alpha>0$ as the active strength~\cite{Simha_2002}. The director $\nd$ of nematics with high aspect ratio is described by the  Leslie-Erickson equation~\cite{kleman_soft_2003}: $\p{t}\nd=\tens{\Omega}\cdot\nd+\tens{E}\cdot\nd+\mu\nabla^2\nd$, with $\tens{\Omega}=[(\nabla\vect{v})^{\text{T}}-\nabla\vect{v}]/2$ as the vorticity tensor, $\tens{E}=[(\nabla\vect{v})^{\text{T}}+\nabla\vect{v}]/2$ as the strain rate tensor, and $\mu\nabla^2\nd$ penalizes deformations. The resulting perturbation equation of $\delta\tilde{n}_y$ is: ( $\tilde{}$ denotes Fourier transform)
\begin{equation}
    \p{t}\delta\tilde{n}_y=[\alpha(\cos2\theta+1)/2-\mu] k^2\delta\tilde{n}_y
\end{equation}
which implies a competition between torques of activity $\sim\alpha$ and elasticity $\sim\mu$, with $\theta$ as the angle between $\vect{k}$ and the initial direction of order $\nd_0$. The growth rate peaks at $\theta=0$, and thus represents the pure bend mode because the leading term is $\nd_0\times\vect{k}\times\delta\nd$~\cite{de1993physics}. Detailed instability and Fourier analyses supporting bend instability, both for the continuum model and simulations, are provided in the SI.

\emph{Initial disclinations.}---While disclinations in bulk 3D nematics initially appear as small loops~\cite{kleman_soft_2003}, visualizing the surrounding $\nd$ field remains challenging. Even though the formalism for parameterizing $\nd$ has been elucidated~\cite{Binysh_2018}, existing approaches for visualization either focus on local normal cross-sections~\cite{duclos_science_2020, Binysh_2020, Friedel_1969}, thereby missing large-scale structural features, or examine the asymptotic behavior far from the loops~\cite{Houston_2021}. In our system, however, the directors primarily vary within the $\nd_0$–$\delta\nd$ planes, making the spatial structure amenable to an understandable quasi-2D description.  We develop a tool to achieve this as follows.

 Consider a small box enclosing the loop. The second moment of $\nd$ within the box is $\tens{q} = \langle\nd\nd\rangle_{\text{box}}$, with eigenvectors $\norm{N}$, $\norm{M}$ and $\norm{L}$. $\norm{N}$ corresponds to the largest eigenvalue and hence represents the local average orientation of $\nd$. $\norm{M}$ is the intermediate eigenvalue and captures the average deviation between $\nd$ and $\norm{N}$.

This local coordinate system naturally encompasses the structure of the bend instability lying in different $\norm{N}-\norm{M}$ planes, which we call the \emph{principal planes}, and provides the backbone of the analysis reported here. As initial loops emerge from an ordered $\nd$ field, $\norm{N}$ and $\norm{M}$ respectively serve as the undisturbed background ($\nd_0$) and perturbation orientation ($\delta\nd)$, a conclusion similar to the model in Ref.~\cite{Binysh_2018, Binysh_2020}. Thus, $\norm{n}$ is assumed to lie on the principal planes. Furthermore, since $\vect{k}$ mainly aligns with $\norm{N}$ during the instability as discussed above, the pattern on each principal plane already captures the dominant deformation. As a result, this model yields a quasi-2D structure, which is both readily understood and captures the dominant physics. SI Video 3 shows an example of this quasi-2D structure as an initial loop scanned by principal planes.


We propose a unified theory of bend-induced loop emergence via principal plane scans. Fig.~\ref{fig:observe}.a shows that loops stem from the bend instability. Given that active forces are extensile and aligned with $\nd$, bending induces a unit active force approximately along $\norm{M}$, distorting the field and creating a pair of $\pm1/2$ 2D-like defects on each principal plane (Fig.~\ref{fig:observe}.b). Ultimately, defects across different principal planes collectively form the initial disclination loop.

\begin{figure}[b]
\includegraphics[width=\columnwidth]{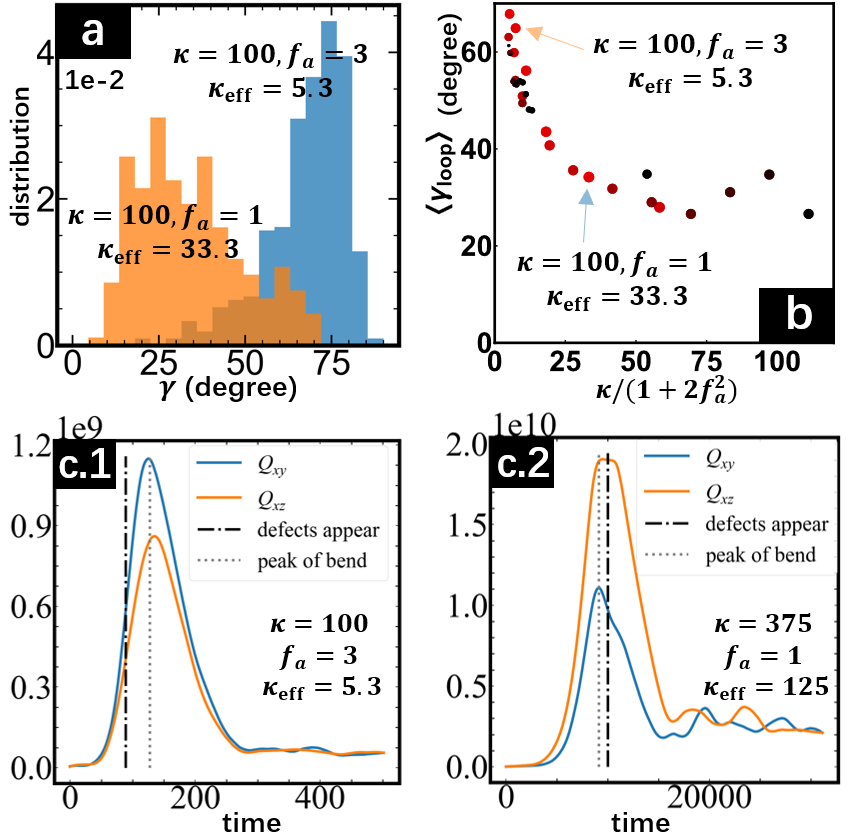}
\caption{\label{fig:timescale} Analysis of initial loops in simulations. \textbf{a)} $\gamma$ distribution of initial loops in different simulations. \textbf{b)} Collapsed curve of average $\gamma$ of all initial loops, showing lower $\ke$ leads to larger $\gamma$. Darker points indicate higher stiffness and larger points indicate lower activity.  \textbf{c)} Interplay between Frank distortions and loop emergence. The colored lines denote the largest Fourier coefficients of $\tens{Q}$. The time of each peak (gray lines), $\tau_{\text{twi}}$, denotes the time scales of growth of twist. The black lines give the nucleation timescale $\tau_\text{nuc}$ when the first loop appears. Panel c) indicates that high $\ke$ results in $\tau_\text{nuc}>\tau_{\text{twi}}$ and thereby induces initial loops with smaller $\gamma$.}
\end{figure}

The microscopic influence on this framework is also characterized by $\ke$. At low $\ke$, defect positions on the principal planes approximately coincide, making the loop perpendicular to these planes (Fig.~\ref{fig:observe}.b.1). At high $\ke$, defect positions differ, causing the loop to tilt (Fig.\ref{fig:observe}.b.2). This tilt can be quantified by the topological parameter $\gamma$, defined as the angle between the loop normal $\norm{\upnu}$ and the rotation vector $\norm{\Omega}$, along which the surrounding directors rotate~\cite{duclos_science_2020}. Since $\nd$ forms the quasi-2D structure, $\norm{\Omega}$ is approximately perpendicular to the principal planes and aligned with $\norm{L}$. Consequently, at large $\ke$, the loop tilt shifts $\norm{\upnu}$ away from the principal planes and reduces $\gamma$. Fig.~\ref{fig:observe}.c and SI Video 4 illustrate the loops' emergence and tilt mechanics. The topological analysis in different simulations reveals that $\langle \gamma \rangle$, the average value of all initial loops, collapses onto a monotonically decreasing curve as a function of $\ke$ (Fig.~\ref{fig:timescale}.b).
 
 This tilt configuration can also be described by additional twist. At high $\ke$, the initial loop is surrounded by directors that vary significantly across different principal planes compared to the low $\ke$ case, as shown in Fig.~\ref{fig:observe}, due to effective shear. This deformation, where $\nd$ behaves differently across planes within the quasi-2D structure, represents \emph{twist}, a feature unique to 3D nematics and absent in 2D. Thus, initial loops at low $\ke$ form primarily through the pure bend mode, while at high $\ke$, significant additional twist is present.

Since our system  dynamics always starts with the bend instability, we interpret the loop tilt as  a competition between two time scales, $\tau_{\text{nuc}}$ and $\tau_{\text{twi}}$ (see the SI for detailed definitions). In Fig.~\ref{fig:timescale}.c, the vertical black lines indicate $\tau_{\text{nuc}}$, the defect nucleation timescale, while the curves represent the time-dependence of the largest Fourier coefficients of $Q_{xy}$ and $Q_{xz}$, the leading order terms of the instability.  The peaks of these curves define $\tau_{\text{twi}}$, marking the transition from bend to other deformation modes. This figure reveals that, at low $\ke$, the active force is strong enough to generate defects immediately during the bend instability as $\tau_{\text{nuc}}<\tau_{\text{twi}}$, generating loops perpendicular to principal planes. In contrast, at high $\ke$, defects do not nucleate until the system has developed significant twist, resulting in tilted loops with lower $\gamma$. SI Fig. 1 shows all components of $\tens{Q}$, demonstrating that the other components correspond to second-order perturbations.

\emph{Discussion.}--- We systematically investigate the emergence of disclination loops in 3D active nematics. Our particle-based simulations provide key insights into the microscopic mechanisms that govern the large-scale defect formation. To enable general conclusions, we use a minimal model that captures the essential nematic behaviors of micro-units, including inter-particle alignment, active forces, and semiflexibility. Our results reveal that an activity-dependent effective stiffness parameter $\ke$ governs defect mechanics, morphology, and steady-state properties, enabling precise control over the system's emergent behavior. This insight could enable the rational design of microscopic parameters to develop self-regulating fluid systems and bio-inspired materials with tunable transport properties.

A crucial innovation of this work is the concept of principal planes, which are natural coordinate systems for expressing director fields in the vicinity of defects ,especially in active nematics. This framework enables visualizing and understanding the complicated 3D structure of initial loops by an approximated quasi-2D configuration. This clarity is reflected in our precise and illustrative description of the previously proposed pure-bend-induced loop mechanism in special cases~\cite{duclos_science_2020, Binysh_2020}, as well as in our extension to more general conditions involving twist. Thus, the principal planes representation allows us to establish a universal theory for defect loop formation in 3D active nematics.  Moreover, as a generalization of the model provided by Friedel and de Gennes~\cite{Friedel_1969}, our framework holds for any orientational field with small perturbations, and thus can be applied to disclination loops and lines in diverse situations including passive nematics or ferromagnetic systems.

Although our model is built for extensile systems undergoing a bend instability, its validity generalizes to other cases. If the system is made contractile or key parameters, such as the flow-alignment coefficient and elastic moduli, are adjusted, the splay instability may dominate, as discussed in  Refs.~\cite{Elias_2016, Nejad_2022, Edwards_spontaneous_2009} and the SI. However, in that case our framework predicts minimal changes in loop emergence as $\norm{k}$ remains within the principal planes, now aligning with $\delta\nd$. More broadly, the loop emergence mechanics remain valid once $\norm{k}$ lies in the principal planes, with only a change in the orientation of the $\pm1/2$ 2D-like defect pairs~\cite{giomi_defect_2014, Tang_orientation_2017}. Moreover, when $\norm{k}$ deviates from the principal planes, effective shear accompanied by additional twist emerges due to $\nd$ gradients across multiple planes (Fig.~\ref{fig:observe}). In particular, when $\norm{k}$ is nearly perpendicular to the principal planes, as in the case of a pure twist mode, the quasi-2D dynamics cease to be applicable. We are developing a more unified framework that accommodates such scenarios in ongoing work~\cite{thesis}.

The initial loops could also be investigated by the existing model that focuses on the topological properties of point defects on the normal cross-sections~\cite{kleman_soft_2003, alexander_colloquium_2012, duclos_science_2020, Friedel_1969}. The collapsed curve of $\langle\gamma\rangle$ indicates that increasing $\ke$ transforms the loops from wedge-twist type to pure-twist. In contrast to this local analysis, our framework examines loops on a larger scale, enabling us to reveal the structure and mechanisms underlying this transition by the competition between Frank elastic energy and active stress, which is well captured by the tilt of principal planes.

Our representation of disclination loops in the principal plane representation immediately suggests a way to produce and control disclination loops in nematic systems. One can readily use this framework to produce loops of desired textures in experimental active nematic systems including microtubule composites~\cite{duclos_science_2020} and  chromonic liquid crystal - swimming bacteria composite systems~\cite{zhou_living_2014}. Further, the steady state properties of the system are determined by $\ke$. This work gives us a design principle for measuring the activity and controlling the bulk properties of these experimental realizations. Hence, it can pave the way for designer active fluids with tailored inhomogeneous mechanical responses. 

In conclusion, we establish a universal theory for disclination loop emergence in 3D active nematics, emphasizing the critical role of activity-dependent effective stiffness. Our particle-based simulations strongly support this mechanism. Together, the theory and simulations suggest further experimental validation, hydrodynamic modeling, and kinetic analyses for a comprehensive understanding of topological defect dynamics in active matter. These insights not only deepen our understanding of non-equilibrium physics but also pave the way for applications of active nematic materials in engineering and biological systems.

\emph{Acknowledgments.}
We thank Gareth Alexander for fruitful advice on the writing of this work. We acknowledge support from the Brandeis Center for Bioinspired Soft Materials: MRSEC DMR-2011846 and NSF: DMR-1855914. Computational resources were provided by NSF ACCESS allocation TG-MCB090163 (Expanse and Stampede) and the Brandeis HPCC which is partially supported by MRSEC DMR-2011846. 
\bibliography{Loops}

\end{document}


\author{Yingyou Ma}
\affiliation{Martin Fisher School of Physics, Brandeis University, Waltham, Massachusetts 02453, USA}

\author{Christopher Amey}
\affiliation{Martin Fisher School of Physics, Brandeis University, Waltham, Massachusetts 02453, USA}

\author{Aparna Baskaran}
\affiliation{Martin Fisher School of Physics, Brandeis University, Waltham, Massachusetts 02453, USA}
\email{aparna@brandeis.edu}

\author{Michael F. Hagan}
\affiliation{Martin Fisher School of Physics, Brandeis University, Waltham, Massachusetts 02453, USA}
\email{hagan@brandeis.edu}

\title{Supplemental Material: The mechanics of disclination emergence in 3D active nematics}
\maketitle

\tableofcontents


\section{Simulation details}
\label{sec:SimDetails}
\subsection{Model and methods}
\label{sec:methods}

Our simulation extends the model developed for 2D active nematics in Ref.~\cite{joshi_interplay_2019} Each filament is modeled as a semiflexible bead-spring polymer. The nematic alignment is ensured by an excluded volume potential between beads, while the nematic active stress arises from an applied bi-directional self-propulsion force directed along the filament tangent on each monomer.  This could be regarded as the simplest model of nematic micro-units that accounts for nematic alignment, nematic active stress and intrinsic stiffness.

The filament dynamics is simulated according to Langevin dynamics, with the equation of motion for each monomer given by:
\begin{equation}
m\ddot{\vect{r}}=\vecfa -\xi\dot{\vect{r}}-\nabla_{\vect{r}}U+\vect{R}(t)
\end{equation}
with $m$ as the monomer mass, $\vect{r}$ as the monomer position, $\vecfa$ as the active force, $U$ as the potential energy whose gradient provides the conservative forces, $\xi$ as the linear friction coefficient and $\vect{R}(t)$ as the zero mean Gaussian white thermal noise with the correlation function given by fluctuation-dissipation theorem:
\begin{equation}
\langle\vect{R}_{i}(t)\cdot\vect{R}_{j}(t')\rangle=6\xi \kt\delta_{ij}\delta(t-t')
\end{equation}
where $\kB$ is the Boltzmann constant, $T$ is the temperature, $i$ and $j$ are monomer indices. $\xi$ is selected to be $10\sqrt{\kB T m/\sigma^2}$ to ensure overdamped dynamics where $\sigma$ is the monomer diameter.

 The interaction potential energy includes three parts: a bond potential between neighboring monomers for the same filament, an angle potential between each neighboring bond and a non-bonded volume-excluded potential between all pairs of monomers of filaments. The bond stretching is controlled by a FENE potential~\cite{fene_1990}:
\begin{equation}
U_{\text{bond}}(r)=-\frac{1}{2}\kbond R_{0}^{2}\ln\left(1-\left(\frac{r-\Delta}{R_{0}}\right)^{2}\right)
\end{equation}
with bond strength $\kbond=2000 \kt$, equilibrium bond length $\Delta=0.5\sigma$ and maximum fluctuation $R_{0}=0.4\sigma$, with $r$ as the bond length and $\sigma$ as the monomer diameter. We set these values to give a mean bond length of $0.4 \sigma$ with standard deviation $0.1\sigma$ to ensure that the filaments remain sufficiently smooth and avoid jamming. 

The angle potential has a harmonic form:
\begin{equation}
U_{\text{angle}}(\vartheta)=\bar{\kappa}(\pi-\vartheta)^{2}
\end{equation}
with filament bending modulus $\bar{\kappa}$ defining the filament stiffness. Here $\vartheta$ is the angle between neighboring bonds. The excluded volume potential avoids overlap of non-bonded monomers and effectively forces local nematic alignment. We use the Weeks-Chandler-Anderson (WCA) potential, which is the truncated repulsive part of the Lennard Jones potential~\cite{Weeks1971}:
\begin{equation}
U_{\text{nb}}(r)=4\varepsilon\left(\left(\frac{\sigma}{r}\right)^{12}-\left(\frac{\sigma}{r}\right)^{6}+\frac{1}{4}\right)
\label{eq:WCA}
\end{equation}
where $r$ is the displacement between two non-bonded monomers, $\varepsilon$ controls the strength of steric repulsion, which we set to $\varepsilon=\kt$. This potential is cut off at the minimum, $r=2^{1/6}\sigma$.

Finally, the activity is modeled by self-propulsion force on each monomer directed along the filament tangent toward its head:
\begin{equation}
\vecfa=\eta_\alpha(t) \bar{f}_{\text{a}}\, b\,\norm{t}
\label{eq:fa}
\end{equation}
where $b$ is the bond length and $\norm{t}$ is the unit tangent vector of each monomer, representing the direction of the active force. This active force is shared equally between $i$th and $(i+1)$th monomer as $\bar{f}_{\text{a}}/2$, where $\bar{f}_{\text{a}}$ is strength of activity. To simulate active nematics, the active force is made to have nematic symmetry by having it reverse direction (for every monomer on a given filament) at stochastic time intervals, which effectively switches the head and tail of the filaments. This is accounted for in Eq.~\eqref{eq:fa} by $\eta_\alpha(t)$, which changes its value between 1 and -1 at Poisson distributed time intervals with averaged value $\tau_\text{r}$, as the time scale controlling the reversal frequency. We set $\tau_\text{r}$ to $\sqrt{m\sigma^2/\kt}$ where $m$ is the monomer mass. This timescale has been shown to effectively capture the nematic active force on timescales much longer than $\tau_\text{r}$ \cite{joshi_interplay_2019}.

We used a modified version of LAMMPS \cite{LAMMPS}, which incorporates the necessary functions for active force. The additional active code is available in Ref.~\cite{GitHub}. Within the LAMMPS framework, the LJ unit system was employed, wherein the length unit corresponds to the diameter of a monomer $\sigma$, the energy unit is $\kt$, and the mass unit is the mass of a monomer $m$. Our simulations used $N\approx2.8\times10^5$ rods in a periodic box with edge length $L=200 \sigma$, equilibrium bond length $b=0.5 \sigma$, and 50 beads per rod. As noted in the main text, we simulated a wide range of stiffness and activity values (which we report in dimensionless units) $\kappa\equiv\bar{\kappa}/\kt \in [100,1500]$ and $\fa \equiv \bar{f}_\text{a} \sigma/\kt \in [1-10]$. 

For each simulation, our initial condition was a perfectly ordered state, achieved by placing straight filaments at uniformly random positions throughout the simulation box, all oriented along the $\hat{x}$ direction. The random positioning accounted for the periodic boundary conditions, such that filaments could span across and wrap around the box edges. Our initial condition was a perfectly ordered state, which we obtained by positioning straight filaments oriented along the $\hat{x}$-direction in random positions throughout the box, accounting for periodic boundary conditions. To eliminate high-energy interactions (e.g. overlaps), we used the soft pair potential instead of WCA potential during the initial thermalization. The soft pair potential is given as
\begin{equation}
    U_{\text{soft}}(r) = A\left[ 1 + \cos\left(\frac{\pi r}{r_c}\right) \right] \qquad r<r_c
\end{equation}
which does not blow up as $r$ goes to 0. Here we set $A=30$ and $r_c=1$. The high-energy interactions are removed by energy minimization, with the stopping tolerance for energy as $10^{-6}$ (unit-less), the stopping tolerance for force as $10^{-6}$ (in force units), max iterations of minimizer as 2000 and max number of force/energy evaluations as 2000. The process of eliminating overlaps and thermalization introduced sufficient perturbations to trigger the onset of the instability despite the aligned initial condition. Once the system was equilibrated, we switch the pair potential to WCA type, turn on the acivity and start the Langevin integrator. For all simulations, we used time step length $\Delta t=0.01$. The total simulation time ranges up to $1.6\times10^5 \tau_\text{r}$ ($1.6\times 10^7$ timesteps) depending on how long simulations required to reach steady-state. The total time for each simulation is given in in the file \emph{parameters.csv} in \textbf{Directory} \emph{Simulation} at~\cite{GitHub}.

\subsection{Coarse-graining techniques}\label{sec:coarse}
We use the following coarse-graining techniques to derive the uniaxial $\Q$ field from the configurations of the particle-based simulations of active semi-flexible filaments.

 The nematogen micro-units are defined as the normalized bond vectors $\norm{b}$ between neighboring particles within the same filament. The simulation box is divided into $N_{\text{int}}\times N_{\text{int}}\times N_{\text{int}}$ voxels, and in each voxel, we compute the local order parameter $\Q_{\text{int}} = \langle \norm{b}\norm{b} - \tens{I}/3 \rangle$, representing the average order parameter within the grid. The discrete Fourier transform is then applied to $\Q_{\text{int}}$, resulting in $\tilde{\Q}_{\text{int}}$, where high-frequency components above a threshold wave vector $k_{\text{threshold}}$ are discarded to focus on the long-wavelength dynamics considered in continuum analyses. To further reduce spatial noise, we apply an inverse Fourier transform to $\tilde{\Q}_{\text{int}}$ with a Gaussian filter, effectively replacing each voxel's $\tens{Q}$ value from a Kronecker delta function to a Gaussian packet characterized by standard deviation $\sigma_{\text{Gauss}}$. This process yields the coarse-grained $\Q$ field, where the largest eigenvalue corresponds to the scalar order parameter $2S/3$, and the associated eigenvector defines the director $\norm{n}$. During the inverse transform, the grid resolution of the output data, $N_{\text{final}}\times N_{\text{final}}\times N_{\text{final}}$, modulates the smoothness of the coarse-grained field.

Three parameters control the degree of coarse-graining: $N_{\text{int}}$, $k_{\text{threshold}}$, and $\sigma_{\text{Gauss}}$. The influence of $N_{\text{int}}$ becomes negligible if the initial voxel size is substantially smaller than $\sigma_{\text{Gauss}}$, as the Gaussian packet spans multiple voxels. Coarser results are achieved with lower $k_{\text{threshold}}$ values and higher $\sigma_{\text{Gauss}}$ values.

For the following discussion, we non-dimensionalize lengths by the particle diameter $\sigma$. In our simulations, the system spans a periodic boundary box of $200  \times 200 \times 200$ We set $N_{\text{int}} = 300$, $k_{\text{threshold}} = 128$, and $\sigma_{\text{Gauss}} = 2$. Depending on the required smoothness, $N_{\text{final}}$ varies from 128 to 400.

We selected these parameters based on empirical observations. A rough criterion for assessing the effectiveness of the coarse-graining process is whether the positions of the disclination points closely align with the $S$ contours, which are set within the range of 0.15 to 0.3. Details of the algorithm for defect detection are provided in section~\ref{sec:appendix_detect}.

\section{Instability analysis of the continuum field}
\label{sec:Instability}
In general, instability is a crucial aspect of any dynamical system, but it holds even greater significance in active nematics. In equilibrium, passive nematics maintain uniformity throughout the bulk, characterized by a constant $S$ and no gradient in the director field. However, even in the steady state of active nematics, this translational symmetry is broken due to persistent distortions in the director field, as local alignment may become unstable under activity. Such instability can give rise to complex dynamics, including the spontaneous creation and annihilation of defects, turbulent flow states and the emergence of intricate patterns. Therefore, a thorough instability analysis is essential for understanding the fundamental mechanisms driving the emergent spatiotemporal dynamics in active nematics.

\subsection{Continuum theory model}
\label{sec:ContinuumTheory}

The continuum model for suspensions of self-propelled nematogens consists of three fields, the flow field $\vect{v}$, the director field $\nd$, and the density field $\rho$. 

 The average volume fraction $\bar{\varphi}$  of filaments in our simulation is a fixed parameter set to 0.7, a comparably high value chosen to enhance nematic alignment. According to our observation, the relative fluctuations in volume fraction field remain below $10\%$, allowing us to approximate the density field as constant. 

 For simplicity, we implicitly model the solvent dynamics in our simulations and neglect long-range hydrodynamic interactions. $\vect{v}$ is thus governed by the local balance between Stokesian linear friction and active stress $\sigma $:
\begin{align}
    -\zeta\ve + \nabla\cdot\siga &= 0 \label{eqn:v}\\
    \siga &= -\alpha\,\nd\nd
\end{align}
with $\zeta$ as the friction coefficient. The active stress can be understood as follows. The nematogens are active as they push or pull along their long axes, generating the active stress $\siga$ through forces exerted by the micro-units. Since we enforce the active force to be nematic, we assume the locally averaged active stress to be proportional to $\nd\nd$, which represents the locally averaged orientation of nematogens' long axes. Positive and negative values of $\alpha$ correspond to the extensile (pushing) and contractile (pulling) stress, respectively~\cite{Simha_2002}. Since the active filaments in our simulation perform self-propulsion,  the system is  extensile with positive $\alpha$.

 The simplest model describing the time evolution of $\nd$ is governed by advection and relaxation driven by free energy minimization in the overdamped limit, as captured by the standard Leslie–Ericksen equation:
\begin{equation}
    \p{t}\nd=\tens{\Omega}\cdot\nd+(\tens{I}-\nd\nd)\cdot(\lambda\,\tens{E}\cdot\nd+\mu\nabla^2\nd) \label{eqn:n}
\end{equation}
with $\tens{\Omega}=[(\nabla\vect{v})^{\text{T}}-\nabla\vect{v}]/2$ as the vorticity tensor, $\tens{E}=[(\nabla\vect{v})^{\text{T}}+\nabla\vect{v}]/2$ as the strain rate tensor, and $\mu\nabla^2\nd$ prevents deformation due to nematic alignment. We include the projection operator $(\tens{I} - \nd\nd)\cdot$, where $\tens{I}$ denotes the identity tensor, to ensure consistency with the normalization constraint of $\nd$. Finally, $\lambda$ denotes the parameter characterizing the alignment between nematogens and local flow.

\subsection{Instability analysis of the theoretical model}
\label{sec:InstabilityTheory}

 To understand the instability of nematic order, the unperturbed state is assumed to be fully aligned with $x$-direction as $\nd_0=(1,0,0)$, and the perturbation is hereby expressed as $\delta\nd=(0,\delta\nd_y, \delta\nd_z)$. We simplify the expression of perturbed equation  by omitting the derivatives in $z$-direction. There is no loss of generosity because of the symmetry between $y$ and $z$ while the initial state aligns with $x$-direction. The perturbed equation is given by:
\begin{align}
    \p{t}\delta\tilde{\nd}_y &= \left[\frac{\alpha}{2}(\cos2\theta+\lambda)-\mu\right]k^2\delta\tilde{\nd}_y \label{eqn:pert1}\\
    \p{t}\delta\tilde{\nd}_z &= \left[\frac{\alpha}{2}(1+\lambda)\cos^2\theta-\mu\right]k^2\delta\tilde{\nd}_z \label{eqn:pert2}
\end{align}
where $\tilde{\cdot}$ denotes a Fourier transform, $\vect{k}$ is the wave vector and $\theta$ is the angle between $\vect{k}$ and initial order $\nd_0$. This implies a competition between torques of activity $\sim\alpha$ and the Frank elastic energy $\sim\mu$.

 When $\theta=0$, the wave vector aligns with $\nd_0$, which corresponds to a pure bend mode both for $\nd_y$ and $\nd_z$ because $\nd_0\times\vect{k}\times\delta\nd$ is the leading order of the bend deformation $|\nd\times(\nabla\times\nd)|^2$. When $\theta=\pi/2$, as $\vect{k}$ aligns with $\delta\nd_y$, the situation changes slightly. For $\delta\nd_y$, the wave vector aligns with the perturbation orientation and accordingly corresponds to pure splay, because $\vect{k}\cdot\delta\nd$ is the leading order of the splay deformation $(\nabla\cdot\nd)^2$. For $\delta\nd_z$, $\theta=\pi/2$ means that the wave vector is in the direction of $\nd_0\times\delta\nd$, resulting in a pure twist mode because $\nd_0\cdot\vect{k}\times\delta\nd$ is the leading order of the twist deformation $(\nd\cdot\nabla\times\nd)^2$.  In summary, the equation of $\delta\nd_y$ represents the bend-splay mode, where $\theta=0$ denotes pure bend and $\theta=\pi/2$ denotes pure splay. The equation of $\delta\nd_z$ represents the bend-twist mode, where $\theta=0$ denotes pure bend and $\theta=\pi/2$ denotes pure twist.

 Since $\alpha > 0$ in our extensile system, the active torque decreases with increasing $\theta$, indicating that pure bend is the most unstable mode during the instability. In particular, when $\theta = \pi/2$, the active torque associated with $\delta\nd_z$ vanishes, effectively prohibiting the pure twist mode in active nematics, regardless of whether the system is extensile or contractile. The dominance of pure bend and the suppression of pure twist during instability in 3D active nematics has also been reported in previous theoretical studies on wet systems~\cite{Minu_bend_2020, Nejad_2022}.

This consistency between wet and dry active nematics can be understood by the following Fourier analysis. If the long-ranged hydrodynamic interaction is included in the flow field, the velocity equation~\ref{eqn:v} is replaced by
\begin{equation}
     -\zeta\ve+\eta\nabla^2\ve + \nabla\cdot \siga = 0
\end{equation}
where $\eta$ is the Newtonian viscosity coefficient. Together with the incompressibility condition $\nabla\cdot\ve=0$, the velocity is given by
\begin{equation}\label{eqn:ve_fourier}
    \tilde{\ve} = \frac{ik}{\zeta+\eta k^2}(\norm{k}\norm{k}-\tens{I})\cdot\tilde{\s_a}\cdot\norm{k}
\end{equation}
where $k=|\vect{k}|$ and $\norm{k}=\vect{k}/k$~\footnote{Notably, the $\norm{k}\norm{k}$ term vanishes in compressible fluids. In our dry case, incompressibility is not enforced due to allowed monomer volume fluctuations. In wet systems, it is essential since $\ve$ represents the solvent velocity instead.}. This equation shows that the only difference of activity between dry and wet systems lies in the $\vect{k}$-dependence of active torque, so their instability conditions remain qualitatively similar.

Notably, Eq.~\ref{eqn:pert1} also indicates that in contractile systems, where $\alpha<0$, the system would exhibit a pure splay mode when $0<\lambda<1$. This is also reported in the study of 2D dry active nematics~\cite{Elias_2016}. The generalized instability condition can be represented by the dimensionless number $\psi_\pm=\alpha(\lambda\pm1)/(2\mu)$. Pure bend (splay) is the most unstable mode when $\psi_+$ ($\psi_-$ ) is larger than 1.

Eq.~\ref{eqn:n} offers a simplified perspective on nematic order parameter dynamics, providing an intuitive understanding of the bend instability. However, this model's applicability is restricted to highly-ordered nematic systems where significant gradients in the scalar order parameter $S$ are absent, and the orientational order can be effectively represented by a normalized vector $\nd$. Consequently, this model holds only for the initial stages of instability. A more comprehensive description of the order parameter field is given by $\tens{Q}$ in the Beris-Edwards equation. Furthermore, an instability analysis conducted on an even more detailed model, extending beyond the Beris-Edwards framework by incorporating a wider range of coefficients, including distinct elastic moduli, is discussed in Ref.~\cite{thesis}.

\subsection{Instability analysis of the computational model}
\label{sec:InstabilitySimulations}

\begin{figure}[H]
    \begin{subfigure}{0.49\textwidth}
    \centering
        \includegraphics[width=\textwidth]{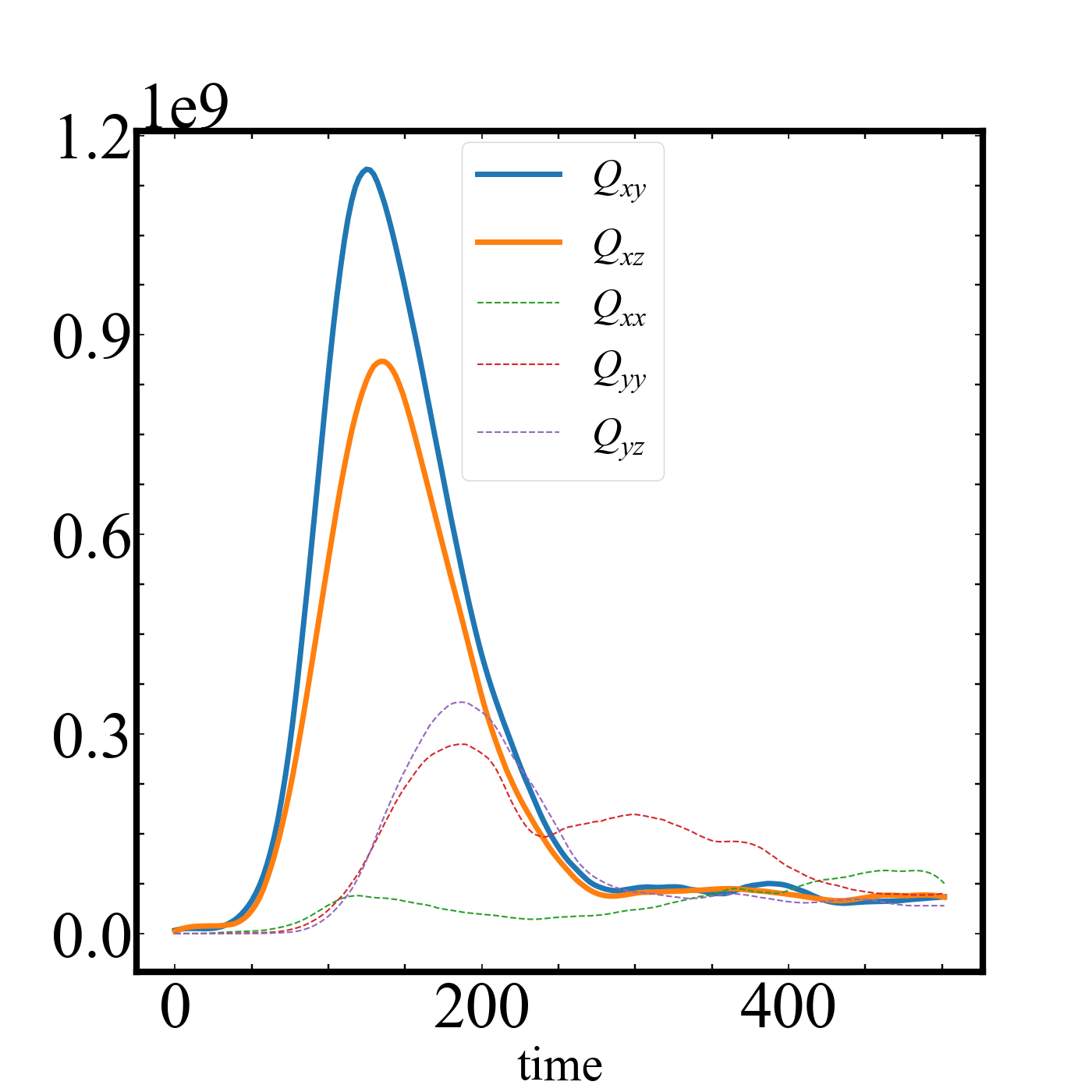}
        \caption{(a)}
    \end{subfigure}    
    \begin{subfigure}{0.49\textwidth}
    \centering
        \includegraphics[width=\textwidth]{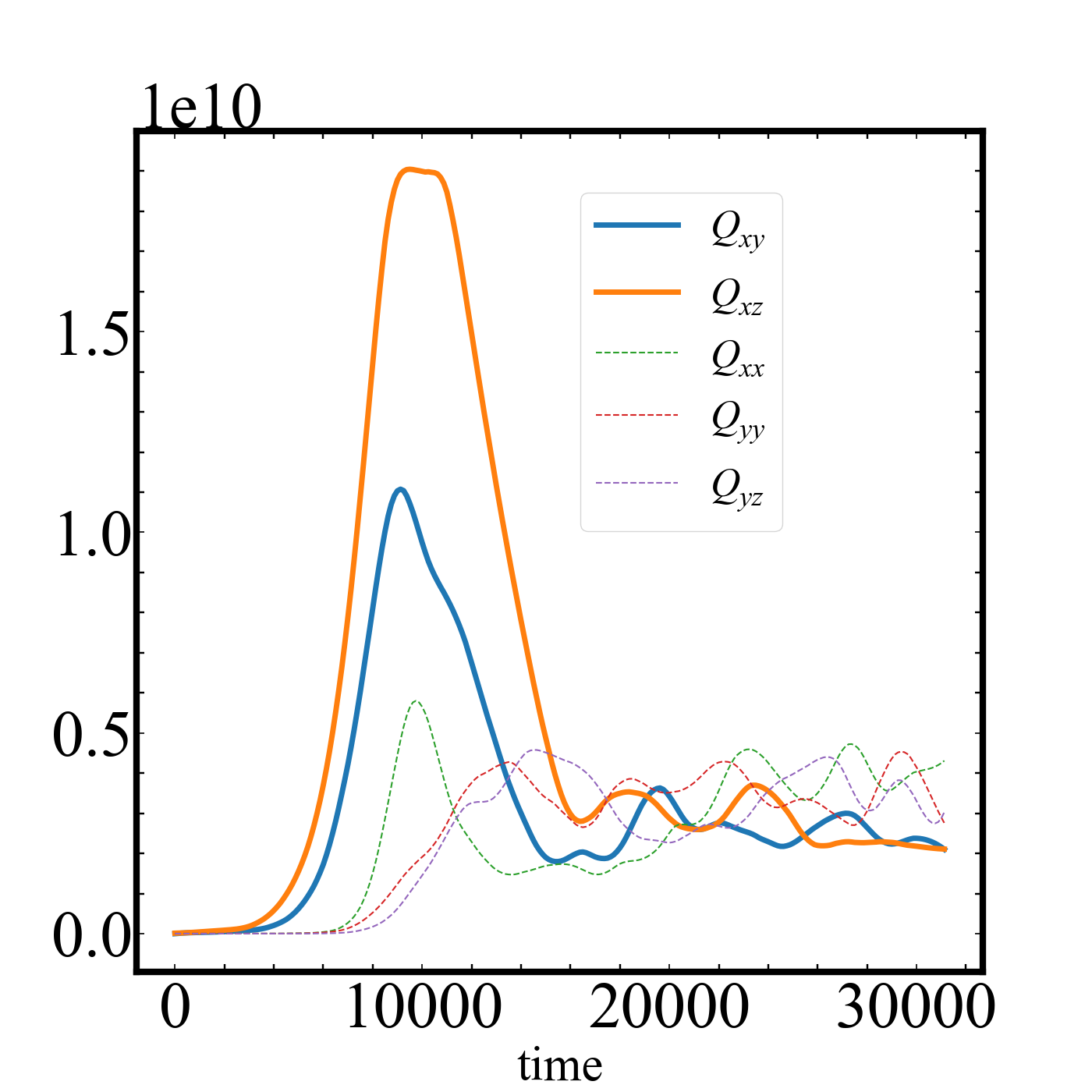}
        \caption{(b)}
    \end{subfigure}  
    \caption{ Examples of results from the instability analysis of simulations conducted at different $\ke$ values. The largest Fourier coefficient for each component of $\Q$ is plotted as a function of time. \textbf{(Left)}: A relatively small effective stiffness: $\kappa=100$, $\fa=3$, giving $\ke=5.26$. \textbf{Right}: A relatively large effective stiffness: $\kappa=375$, $\fa=1$, giving $\ke=125$.
    }\label{fig:fourier_exam}
\end{figure}

We begin by assessing the validity of our theoretical instability analysis. In Section~\ref{sec:InstabilityTheory}, we assume that perturbations arise solely from small reorientations of the director field, expressed as $\delta\nd = (0,\delta\nd_y,\delta\nd_z)$, while variations in scalar order, $\delta S$, are neglected. Under this assumption, it follows directly that the first-order perturbations in the tensor $\Q$ are limited to $\delta\Q_{xy}$ and $\delta\Q_{xz}$, corresponding respectively to $\delta\nd_y$ and $\delta\nd_z$. This prediction is corroborated by our observations of the largest Fourier coefficients of the components of $\Q$, which indicate that $\delta\Q_{xy}$ and $\delta\Q_{xz}$ dominate over other components (see representative examples in Fig.~\ref{fig:fourier_exam}.)

To investigate the spatial configurations of the system during the instability, we conduct a quantitative analysis using Fourier transforms of the coarse-grained $\Q$ field. For each simulation, we apply the Fourier transform $\tilde{f}(\vect{k})=\int \text{d}\vect{r}\, f(\vect{r})\exp(-i\,\vect{k}\cdot\vect{r})$ to each component of $\Q$. Consistent with the Fourier analysis of continuum model, here we investigate the Frank deformation by the orientation of the dominant wave vector during instability. To quantify this orientation, we select the first three dominant wave vectors, calculate their angles with the $x$-axis, and define  the weighted average angle $\bar{\theta}$:
\begin{equation}\label{eqn:theta_bar}
    \bar{\theta}=\frac{\sum_{l=1}^3 |\tilde{\Q}_l|\arccos(\norm{k}_l\cdot\hat{\norm{x}})}{\sum_{l=1}^{3}|\tilde{\Q}_l|}
\end{equation}
where $\tilde{\Q}$ in this equation represents the Fourier coefficients of $\Q_{xy}$ and $\Q_{xz}$, and $l$ denotes the $l$-th dominant wave vector. We focus on $\Q_{xy}$ and $\Q_{xz}$ because their perturbations are significantly larger and directly correspond to director reorientation, unlike other components of $\tens{Q}$, which exhibit considerably smaller perturbations, as shown in Fig. 1 in the main text.

\begin{figure}[H]
    \begin{subfigure}{0.49\textwidth}
    \centering
        \includegraphics[width=\textwidth]{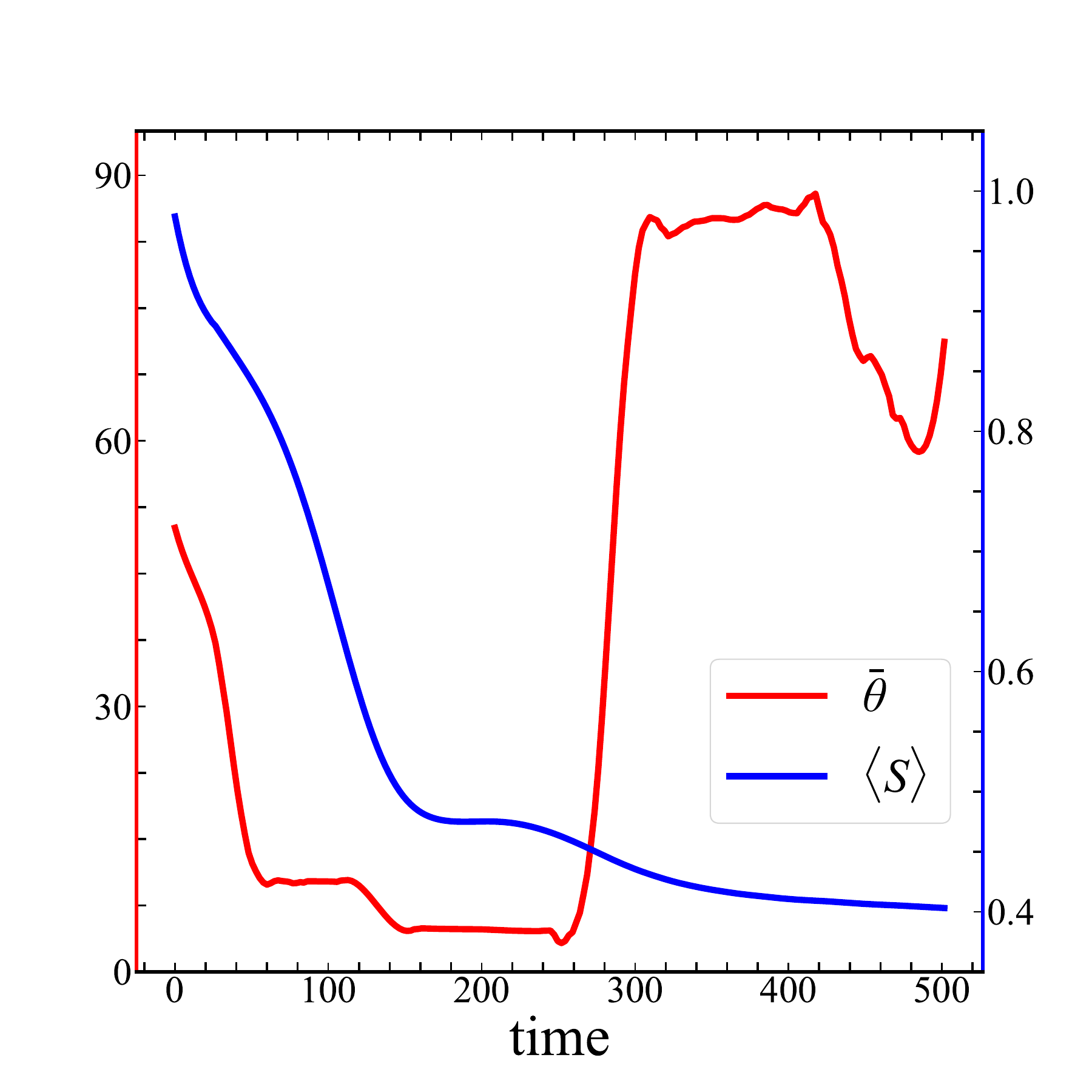}
        \caption{(a)}
    \end{subfigure}    
    \begin{subfigure}{0.49\textwidth}
    \centering
        \includegraphics[width=\textwidth]{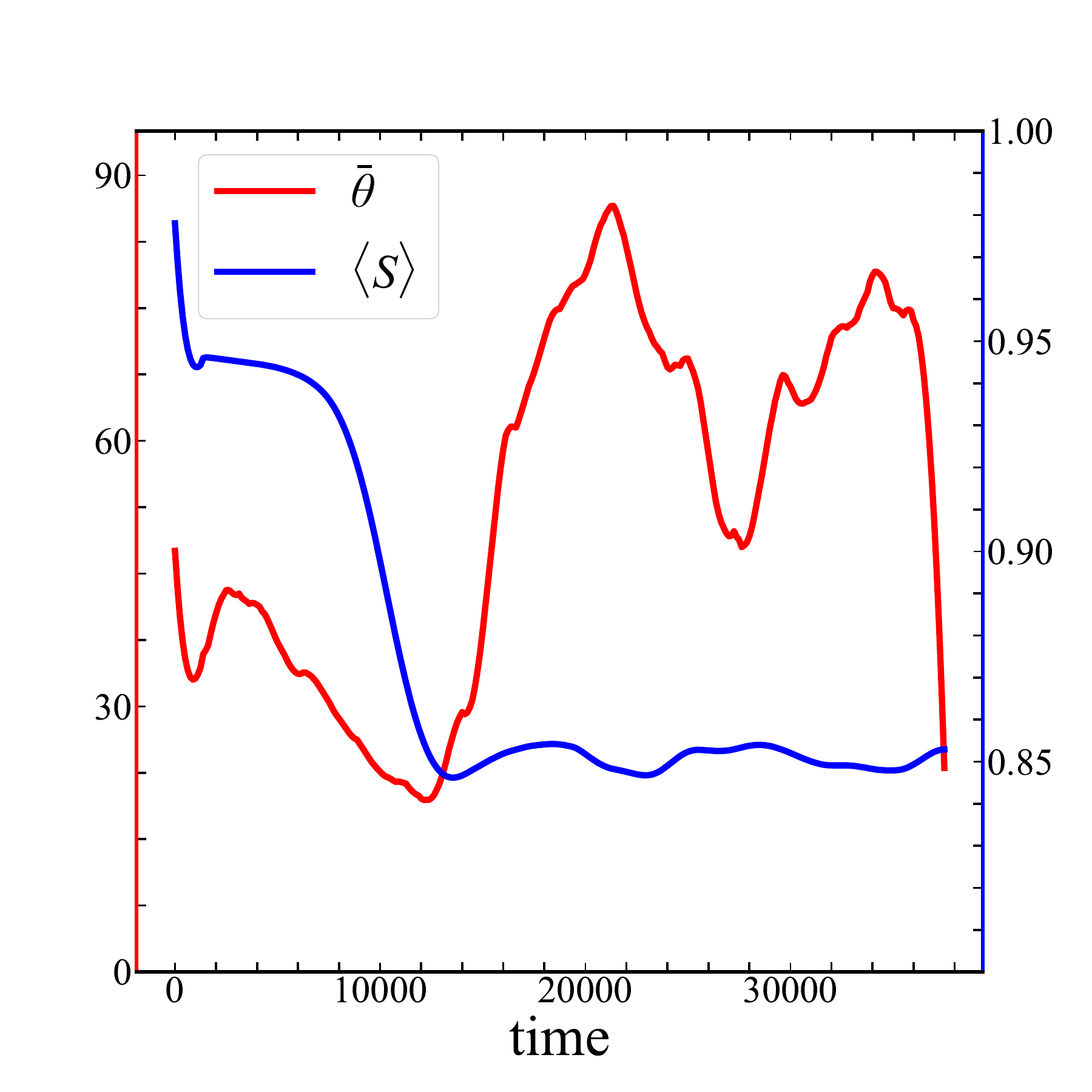}
        \caption{(b)}
    \end{subfigure}  
    \caption{ Examples of results from the instability analysis of simulations conducted at different $\ke$ values. \textbf{Left}: $\kappa=100$, $\fa=3$, giving $\ke=5.26$, comparably small in our parameter set. \textbf{Right}: $\kappa=375$, $\fa=1$, so $\ke=125$, comparably large in our parameter set. The red line represents $\bar{\theta}$ averaged between $\tens{Q}_{xy}$ and $\tens{Q}_{xz}$, corresponding to the $y$-axis on the left (red). The blue line represents  $S$ averaged throughout the whole system, corresponding to  the $y$-axis on the right (blue). 
    }\label{fig:fourier_simulation}
\end{figure}

We show an example of $\bar{\theta}$ as a function of time in the first row of Fig.~\ref{fig:fourier_simulation}. It shows that, after the initial thermalization, $\bar{\theta}$ is essentially much smaller than $45^\circ$ throughout our parameter regime, indicating the wave vector tends to align with $x$-direction during the linear instability. This observation confirms that the bend instability is consistently dominant for all $\ke$ values considered here, coinciding with the perturbation analysis of continuum model.

In addition to the bend instability, Fig.\ref{fig:fourier_simulation} demonstrates that the decrease in $S$ is significantly smaller than the reorientation of the director field. This effect is particularly evident at relatively large $\ke$, as shown in Fig.\ref{fig:fourier_simulation}.b, where $S$ initially stabilizes at a plateau around $S\approx0.95$ before eventually declining. This observation supports the validity of our theoretical continuum model, which considers variations in $\delta\nd$ while neglecting $\delta S$. The stability of $S$ can also be attributed to the protection provided by the Landau–de Gennes free energy, which favors the ordered nematic phase as the equilibrium state. This conclusion is also supported by a more rigorous theoretical analysis involving the time evolution of the tensor order parameter $\tens{Q}$, as discussed in Ref.~\cite{thesis}.

Notably, Fig.~\ref{fig:fourier_simulation}.(a,b) also reveals that, following the linear instability where $\bar{\theta}$ is small, $\bar{\theta}$ always abruptly rises to a high value near $\bar{\theta}\approx90^\circ$. This behavior suggests the presence of nonlinear dynamics, which we have not investigated in this study. However, the time of onset of this nonlinear dynamics could be regarded as one of the well-defined time scales describing the initial instability. Another time scale characterizing this stage corresponds to  the appearance of the first disclination loops, which we defined as $\tau_{\text{nuc}}$ in the main text. 

\begin{figure}[H]
    \centering
    \includegraphics[width=0.45\textwidth]{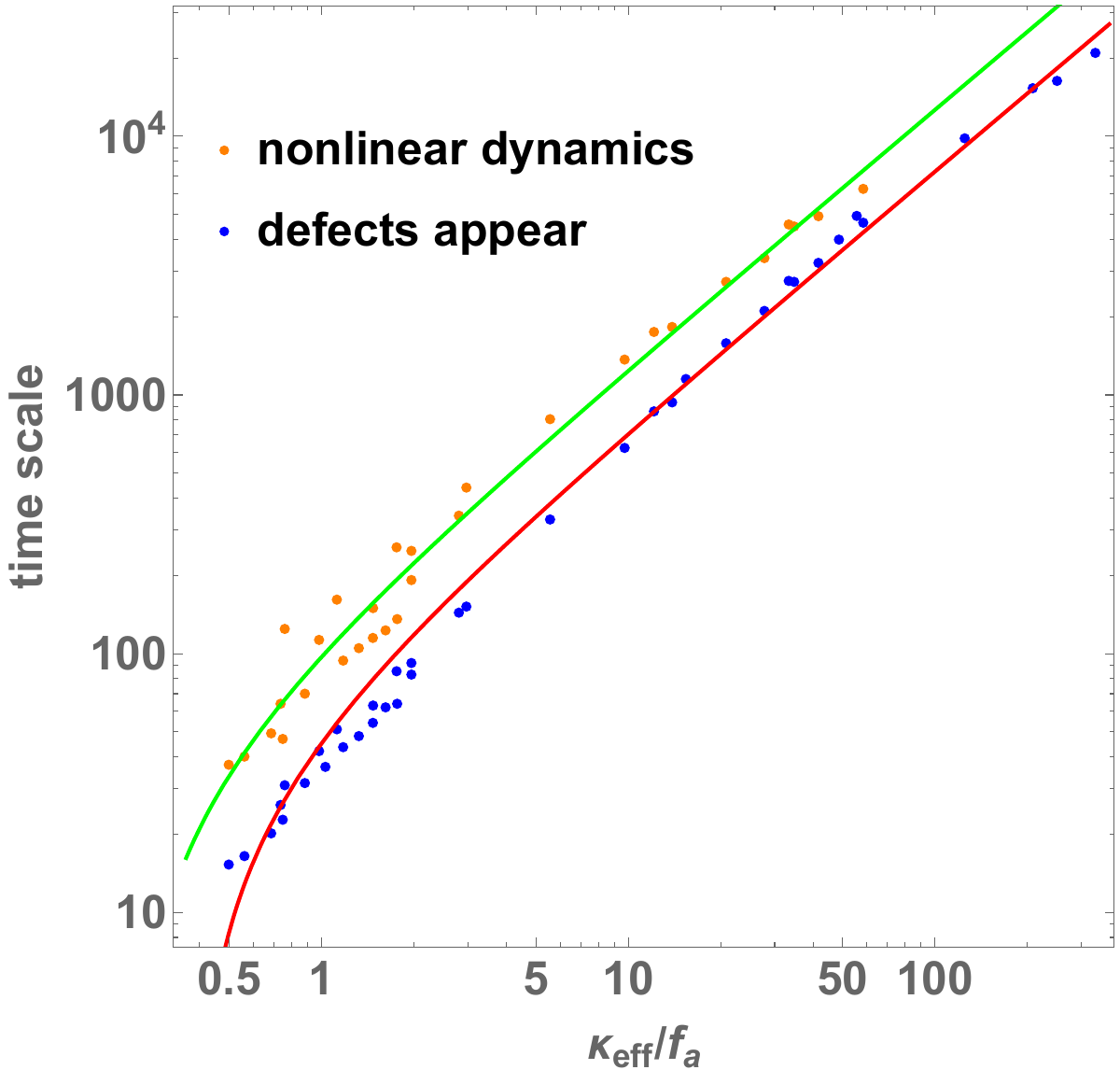}
    \caption{ Scaling analysis of time scales of the initial instability, which collapse is a function of $\ke/\fa$. The blue points represent the time of defect nucleation $\tau_{\text{nuc}}$,   with the red line as the linear fitting line. The orange points represent the time onset of nonlinear dynamics, with the green line as the linear fitting line.  }\label{fig:time}
\end{figure}

 It might be surprising to observe that the rate evolution of system is determined by $\ke/\fa$ rather than $\ke$, as illustrated in Fig.~\ref{fig:time}, where two of the characteristic time scales, the emergence of defects and the onset of non-linear dynamics, are collapsed and linear with $\ke/\fa$. This conclusion implies that, beyond the mechanical scale dictating the system's configuration, the active force provides a linear time scale intrinsic to the dynamics. 


\section{Analysis and theoretical model of disclinations}
\label{sec:analysisDisc}
\subsection{The identification and analysis of disclinations in simulations}\label{sec:appendix_detect}

The disclinations are identified as singular particles which compose into the disclination lines in 3D nematics. We employed the disclination detection algorithm introduced in Ref.~\cite{defect_detect, defect_detect2}, a method which has been widely used in the study of disclinations in 3D active nematics~\cite{Nejad_2022, Digregorio_2024}. The fundamental approach involves calculating the winding number of directors along each unit closed path. In the 3D grid, each voxel contains six unit square paths, each defined by four directors on a $2 \times 2$ lattice site. For each path, we begin with one director and assign the subsequent directors counterclockwise to align as closely as possible with the preceding one. If the dot product between the initial and final director is less than zero, the winding number for that path is considered non-zero, and the path's center is identified as a disclination point. A detailed explanation and visualization of this algorithm are provided in Ref.~\cite{Digregorio_2024}. 

 This defection algorithm returns a group of disclination points. To further classify these points into distinct continuous lines, we use \emph{Hierholzer’s algorithm}~\cite{Hierholzer1873,Fleischner1990}. In our simulations,  the disclinations would always form closed loops due to the periodic boundary conditions. This implies that each disclination point has an even number of neighboring disclination points~\cite{Digregorio_2024}. Here, neighboring disclination points are defined as those located on distinct unit square paths within the same voxel, with the displacement between these two disclination points referred to as an \emph{edge}. Starting from an arbitrary disclination point, we successively trace one of its neighboring points via the edges, eventually returning to the initial point to form a closed loop. This method ensures that all disclination points and edges are accounted for.

 The discussion above focuses on the disclination lines, while the following elaborates the identification and analysis of initial disclination loops. This requires more information about the directors around the loop. The core (or skeleton) of the disclination loop is assumed to consist of the disclination points identified by the winding number, as introduced above. The surrounding low-$S$ region is considered to represent the region around these cores.

\begin{figure}[H]
    \begin{subfigure}{0.34\textwidth}
    \includegraphics[width=1\textwidth]{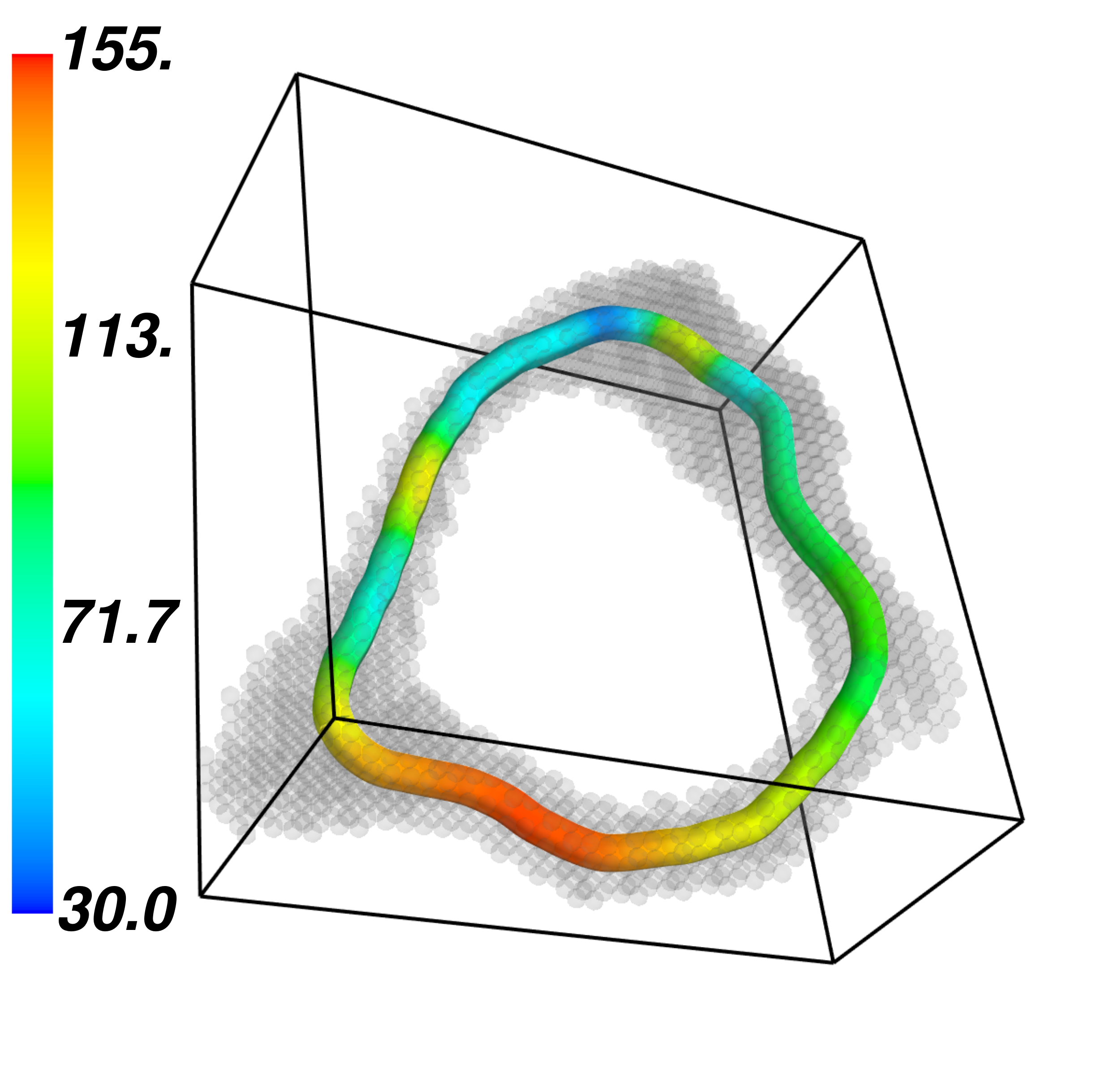}
    \caption{(a)}
    \end{subfigure}
    \hfill
    \begin{subfigure}{0.35\textwidth}
    \includegraphics[width=1\textwidth]{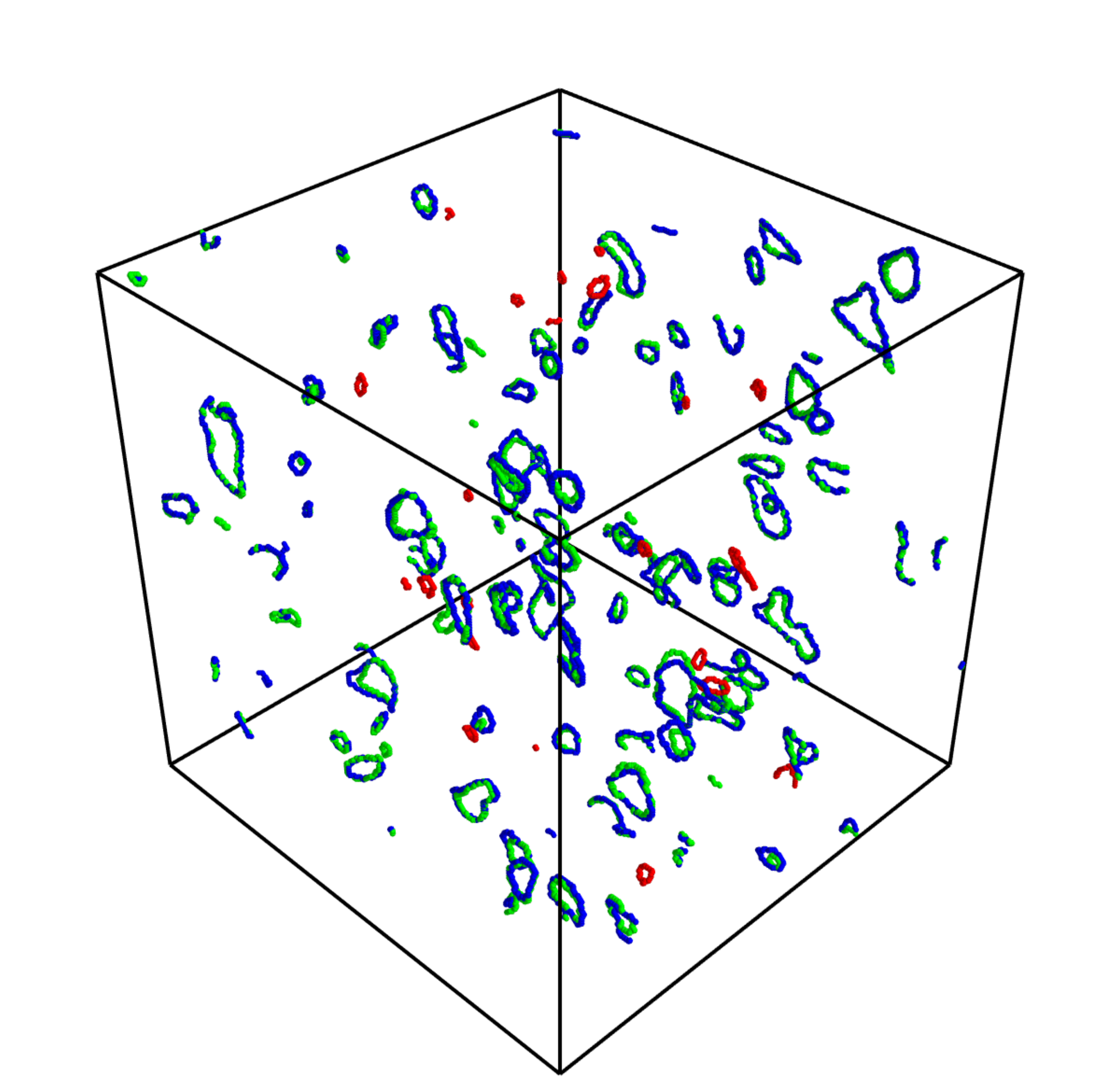}
    \caption{(b)}
    \end{subfigure}
    \hfill
    \begin{subfigure}{0.27\textwidth}
    \includegraphics[width=1\textwidth]{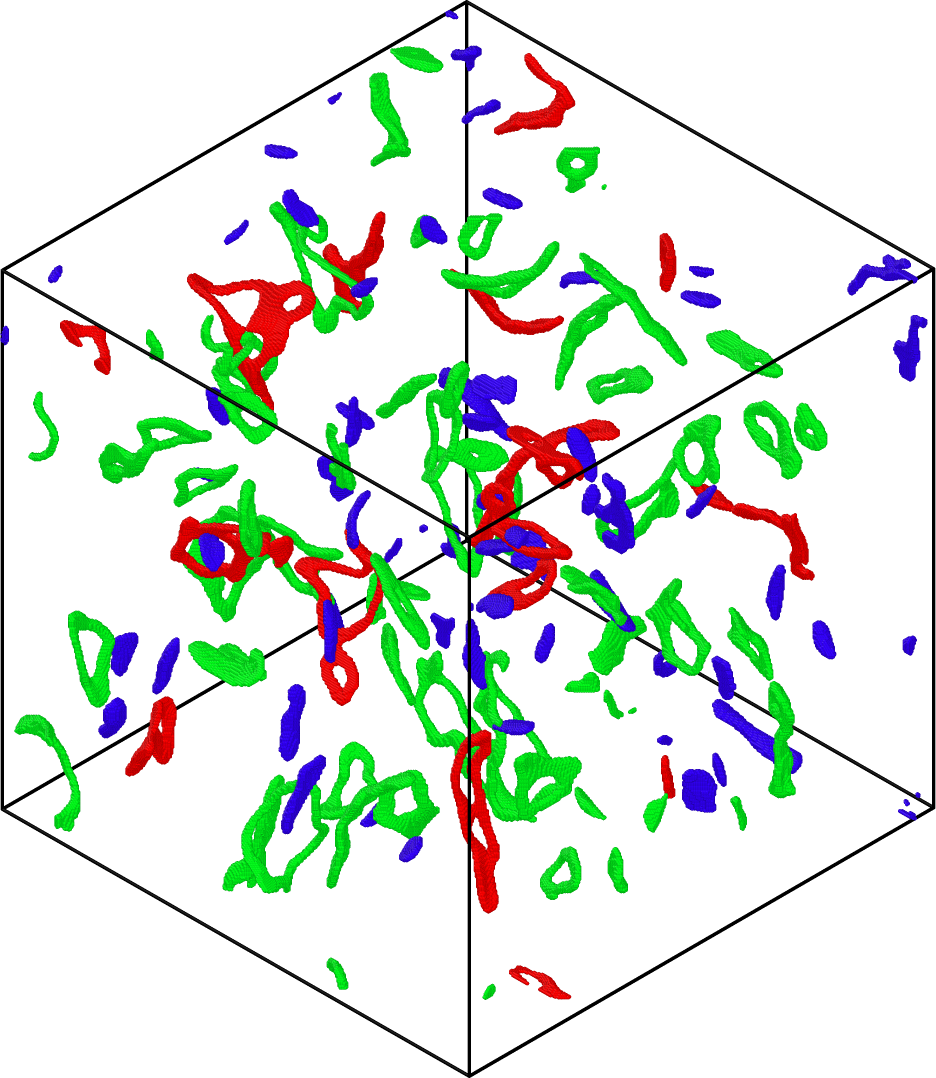}
    \caption{(c)}
    \end{subfigure}
    
\caption{Examples of the analysis of initial disclination loops. \textbf{(a)} The analysis result for an initial dislination loop in the simulation with $\kappa=100$, $\fa=3$, $\ke=5.3$. The gray region represents the independent defect $\mathcal{D}$, enclosed by its the smallest orthogonal bounding box $\mathcal{H}$, shown as the black lines. The curve indicates the skeleton of $\mathcal{D}$, colored by $\beta$ in units of degrees. The coordinates and values of $\beta$ of this loop are smoothened for clarity. \textbf{(b)} Part of the disclination points from the simulation with $\kappa=100$, $\fa=3$. The green points correspond to $t=118.8$, while the blue and red points are from $t=120$. The blue points stand for the disclinations which already existed in $t=118.8$, whereas the red points denote the newly emergent disclinations in $t=120$. This demonstrates that disclination loops are approximately static, enabling tracking of loops based on their locations. \textbf{(c)} All independent defects $\mathcal{D}$ recognized in the simulation of $\kappa=175$, $\fa=1$, $\ke=58.3$, $t=5832$, are colored according to their genus. The blue $\mathcal{D}$ has no holes and is hence classified as a defect pancake, not suitable for loop analysis until it evolves to a loop. The green $\mathcal{D}$ has exactly one hole and is identified as a disclination loop. $\mathcal{D}$ in red has two or more holes. They are interacting loops that are excluded from analysis due to the non-negligible interactions.}
\label{fig:loop_analysis}
\end{figure}

 In our simulations, the center of the disclination loop is found to remain nearly static, as the example Fig.~\ref{fig:loop_analysis}.b shows. This allows us to use the center to trace the loops across different frames. The general criteria for selecting initial loops to analyze are as follows: (a) The loop should be analyzed as soon as possible after nucleation to investigate the mechanics of its emergence. (b) However, during the very early stages of emergence, the low-$S$ region often appears pancake-shaped. While its skeleton forms a loop, this region is considered chaotic, with an internal structure beyond the spatial resolution of coarse-graining. As such, loops at this stage are too premature to analyze. (c) We exclude initial loops that are in close proximity to one another, as their interactions are considered significant and cannot be neglected. The detailed selection rules are outlined below.

 Voxels with $S$ values below a chosen threshold are identified and divided into distinct subsets. The threshold for $S$ is selected between 0.15 and 0.3, based on the structure of the loop's skeleton. Each subset of connected disordered voxels is considered an independent defect, denoted as $\mathcal{D}$. To mitigate the effects of local fluctuations in $S$ that might cause disconnections, each $\mathcal{D}$ is expanded by including all its nearest neighboring voxels, while small objects and small holes are ignored. Consequently, each $\mathcal{D}$ forms a connected polyhedral structure. Using this approach, independent defects $\mathcal{D}$ are identified in each frame, along with the smallest orthogonal bounding box $\mathcal{H}$ that encloses the corresponding $\mathcal{D}$. For a given frame, if the center of $\mathcal{D}_1$ lies within $\mathcal{H}_0$ from the previous frame, $\mathcal{D}_1$ and $\mathcal{D}_0$ are considered to represent the same defect at different times. However, if two defects, $\mathcal{D}_1$ and $\mathcal{D}_2$, have their centers located within the same $\mathcal{H}_0$ in the previous frame, they are considered too close to be treated as isolated loops. An example of $\mathcal{D}$ and its $\mathcal{H}$ is shown in Fig.~\ref{fig:loop_analysis}.a.

 The classification of disclination pancakes and loops is based on the number of holes in $\mathcal{D}$. If $\mathcal{D}$ has no holes, it is treated as a disclination pancake. In such cases, its internal structure is beyond the spatial resolution of coarse-graining, and it is not ready for analysis as a disclination loop. If $\mathcal{D}$ has exactly one hole, it is considered a true disclination loop. $\mathcal{D}$ instances identified as loops in the current frame but classified as pancakes in the previous frame are marked as the initial disclination loops for analysis. If $\mathcal{D}$ has more than one hole, it is considered as representing two or more interacting loops and is therefore excluded from analysis.

 The number of holes in each $\mathcal{D}$ is represented by its topological genus $g$, which is derived from the Euler characteristic $\chi$ using the relation $g = 1 - \chi / 2$. Since each $\mathcal{D}$ is a polyhedron composed of connected voxels, the Euler characteristic $\chi$ is calculated as $\chi = V-E+F$, where $V$, $E$, and $F$ are the number of vertices, edges, and faces of the polyhedron, respectively. The isolated defects $\mathcal{D}$ colored by genus in one example frame are shown in Fig.~\ref{fig:loop_analysis}.c.

We observe that each initial disclination loop identified by the method above is typically convex and lies approximately on a single plane, suggesting it can be treated as a 2D elliptical structure. For each loop, we establish a local cylindrical coordinate system, where the $z-$axis is aligned with the normal vector $\norm{\upnu}$ of the common plane. The loop $\mathcal{D}$ is divided into 24 subgroups based on the azimuthal angle to conduct topological analysis. The tangent vector of the loop, $\norm{t}$, is given by the relative displacement between the neighboring subgroups, while the rotation vector, $\norm{\Omega}$, is given by the approximated normal vector of the directors within each subgroup. Although $\norm{\Omega}$ lacks an inherent orientation in this approximation, it retains sufficient physical information since the analysis primarily focuses on $\cos\gamma=|\norm{\upnu}\cdot\norm{\Omega}|$. Therefore, for each loop, we obtain a single numerical value for $\norm{\upnu}$, 24 discrete values for $\norm{t}$ and $\norm{\Omega}$, and hence 24 values for $\gamma$ and $\beta=\norm{\Omega}\cdot\norm{t}$. The loop-averaged value of $\gamma$, denoted as $\langle\gamma\rangle_{\text{loop}}$ is the average of these 24 $\gamma$ values. An example of the analysis results for initial disclination loops is illustrated in Fig.~\ref{fig:loop_analysis}.a.


\subsection{Principal planes}
\label{sec:PrinciplePlanes}

\subsection{Validation of the theoretical model for the emergence of initial disclination loops}

Based on principal planes, we propose the 2D-analog theory to explain the emergence of initial disclination loops. To examine this model, it is necessary to establish several criteria as follows. \textbf{(a)} The average orientation of directors surrounding the loop, denoted as $\norm{N}$ within a small box bounding the initial loop, should align with the initial order, as the configuration remains close to the initial condition. \textbf{(b)} The norm of the initial loop, represented by $\norm{\upnu}$, should also align with the initial order, as predicted by the geometric conclusions of the 2D-analog theory. \textbf{(c)} Given that the defect exhibits a 2D structure on $\norm{N}-\norm{M}$ planes, the rotation vector $\norm{\Omega}$, which represents the axis along which the local directors rotate, should align with $\norm{L}$. This alignment indicates that the loop corresponds to a wedge-twist type, where ideally $\gamma = 90^\circ$, with $\gamma$ being the angle between $\norm{N}$ and $\norm{\Omega}$.

\begin{figure}[H]
    \begin{subfigure}{0.49\textwidth}
    \includegraphics[width=1\textwidth]{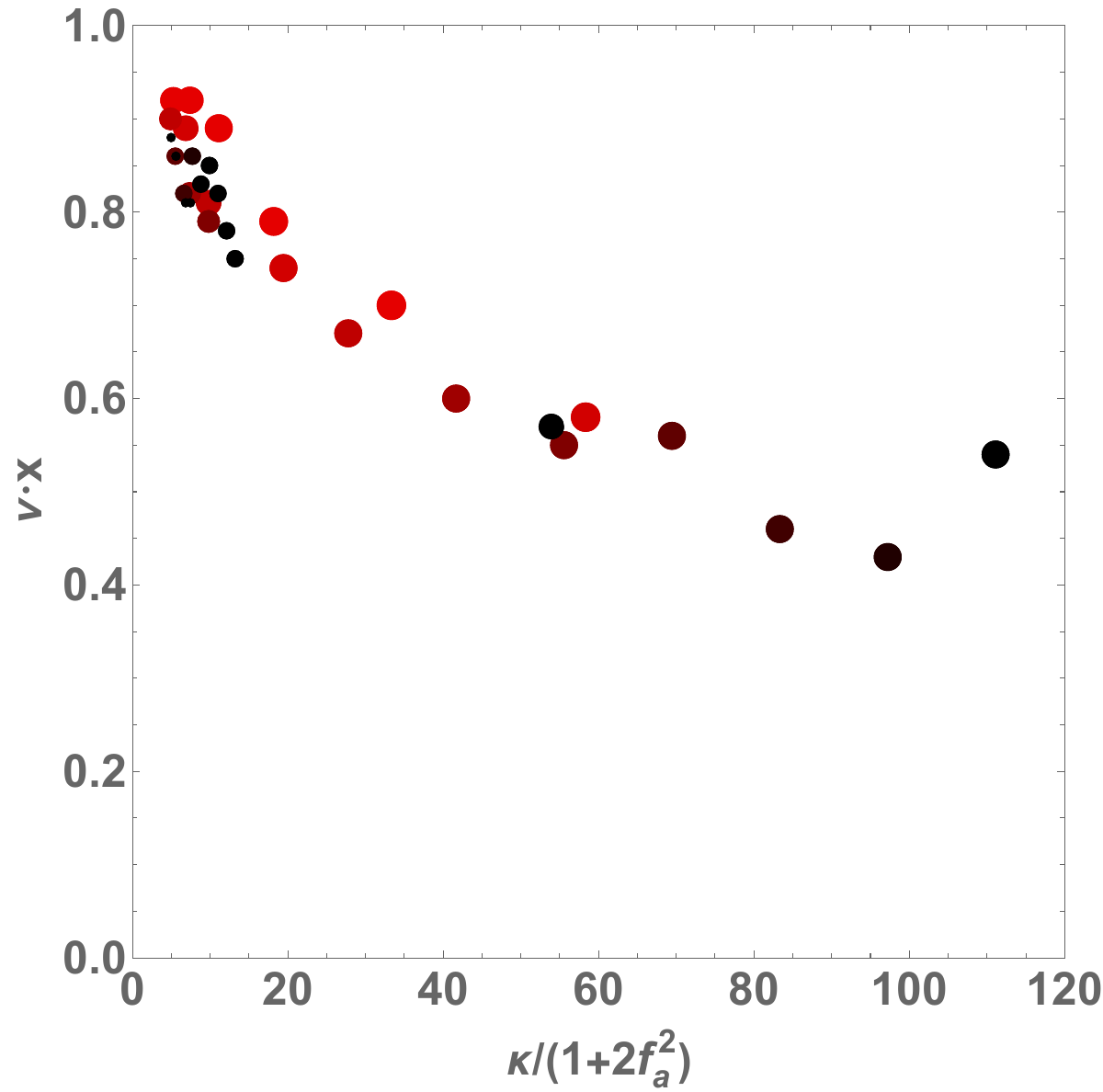}
    \caption{(a)}
    \end{subfigure}
    \hfill
    \begin{subfigure}{0.49\textwidth}
    \includegraphics[width=1\textwidth]{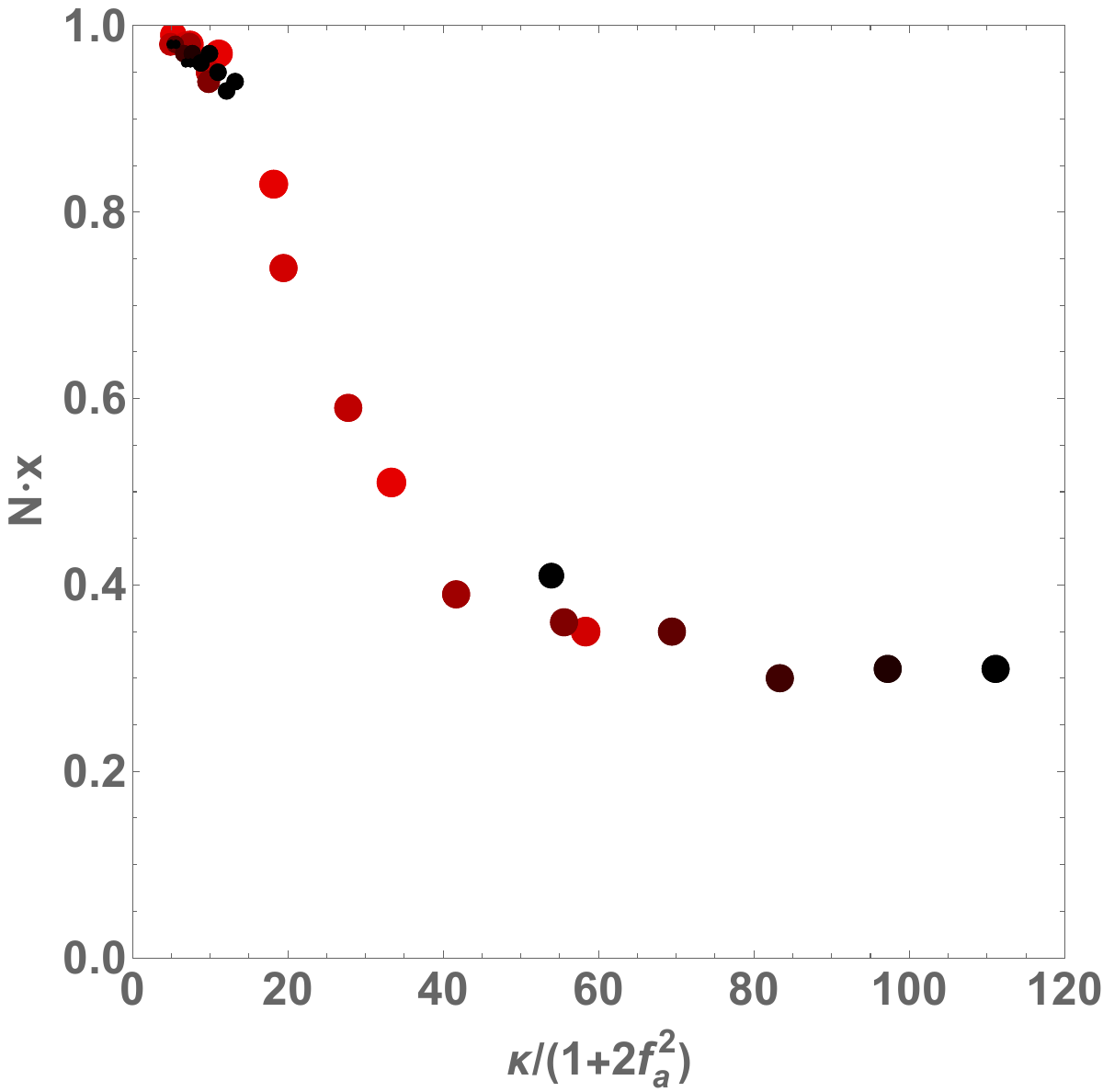}
    \caption{(b)}
    \end{subfigure}
    
\caption{Analysis of the geometric structure of initial loops shows that they depend on activity and stiffness only through the effective stiffness $\ke$. \textbf{(a)} The alignment between the loop's normal vector $\norm{\upnu}$ and the direction of initial order, $\hat{\mathbf{x}}$, as a function of $\ke$.  \textbf{(b)} The alignment between the average orientation of the directors surrounding the initial disclination loops $\mathbf{N}$ and the direction of initial order.}
\label{fig:loop_collapse} 
\end{figure}

While the analysis of $\gamma$ is described in the main text, the other two results of the investigation of initial loops across different simulations are presented in Fig.~\ref{fig:loop_collapse}. The findings demonstrate that the geometrical properties of the initial loops also collapse on $\ke$, similar to the properties observed in steady state. For initial disclination loops in simulations with small $\ke \lesssim 30$, these loops satisfy the criteria outlined earlier, suggesting that the bend-induced-mechanics is valid for initial loops with small $\ke$.

In the supplemental data provided at~\cite{OSF} and~\cite{GitHub}, \textbf{Directory} \emph{Loop-Analysis} presents the distributions of $\gamma$, $\norm{\upnu}\cdot\norm{x}$, and $\norm{N}\cdot\norm{x}$ for the initial loops observed in simulations with various values of $\ke$. We focus on simulations with intermediate $\ke$ values, as low $\ke$ leads to densely packed, interacting loops, whereas high $\ke$ does not yield a sufficient number of loops for meaningful statistical analysis. Additionally, we applied the same loop analysis to defect pancakes prior to their evolution into loops. The resulting distribution of defect pancakes in \emph{Loop-Analysis}  closely corresponds to the distribution for initial loops, which confirms that we selected the loops at a sufficiently early stage of their evolution.

\subsection{Comparison between defect analysis based on principal planes and the existing theory based on normal cross-sections}
\label{sec:comparison}

While we presented loop analysis based on principal planes in the main text and the previous section, we now introduce the existing model that analyzes disclination lines and loops based on normal cross-sections \cite{duclos_science_2020, alexander_colloquium_2012, Binysh_2018, de1993physics, kleman_soft_2003}. This existing model provides supplemental information about the \emph{local} topological properties of defects. 

Consider a point on an arbitrary section of a disclination line. The normal cross-section at this point is well defined by its normal vector $\norm{t}$, which corresponds to the local tangent of the disclination line. The surrounding directors, located on a small loop $\Sigma$ that encloses the defect on this plane, correspond to a curve on the 2D real projective plane $\mathbb{RP}^2$, as the state space for 3D directors. This curve starts from a point on $\mathbb{RP}^2$ and ends at its antipodal point. Assuming that the directors undergo minimal deformation, the corresponding curve forms a semicircle characterized by the normal vector $\norm{\Omega}$~\cite{duclos_science_2020}. Since the directors on $\Sigma$, denoted as $\nd_{\Sigma}$, rotate about $\norm{\Omega}$, $\norm{\Omega}$ is referred to as the rotation vector and determines the geometrical structure of $\nd_{\Sigma}$ through the angle $\beta$ between $\norm{\Omega}$ and $\norm{t}$. When $\beta=0$, $\norm{\Omega}$ is aligned with $\norm{t}$, and thus all $\nd_{\Sigma}$ lie within the normal cross-section. The disclination on the plane then exhibits a purely 2D structure, appearing as a $+1/2$ or $-1/2$ defect, known as a \emph{wedge} defect. In contrast, when $\beta=\pi/2$, $\norm{\Omega}$ lies within the normal cross-section, and there must exist directors that are perpendicular to the plane. This type of defect is referred to as a \emph{twist} defect. In principle, $\ro$, $\norm{t}$ and $\beta$ are functions of arc length along  disclination lines in 3D nematics.

An initial disclination loop is assumed to take the form of an ellipse with a uniform $\ro$ due to energy minimization~\cite{duclos_science_2020, Binysh_2020}. Thus, the loop is characterized by $\ro$, its normal vector $\norm{\upnu}$, and the angle $\gamma$ between $\ro$ and $\norm{\upnu}$. When $\gamma=\pi/2$, $\beta$ varies between $0$ and $\pi$ because the tangent vector $\norm{t}$ rotates by $2\pi$ around the loop while $\norm{\Omega}$ remains fixed within the plane of the loop. Specifically, $\beta$ evolves from $0$ (wedge defect) to $\pi/2$ (twist defect), then to $\pi$ (wedge defect), to $3\pi/2$ (twist defect), and finally returns to $0$. Such loops are therefore termed \emph{wedge-twist} loops. In contrast, when $\gamma=0$, $\norm{\Omega}$ is always perpendicular to $\norm{t}$, corresponding to $\beta=\pi/2$ everywhere. This indicates that the directors in each normal cross-section form defects of the twist type. Such loops are hence called \emph{pure-twist} loops.

\begin{figure}[H]
    \centering
    \includegraphics[width=0.5\textwidth]{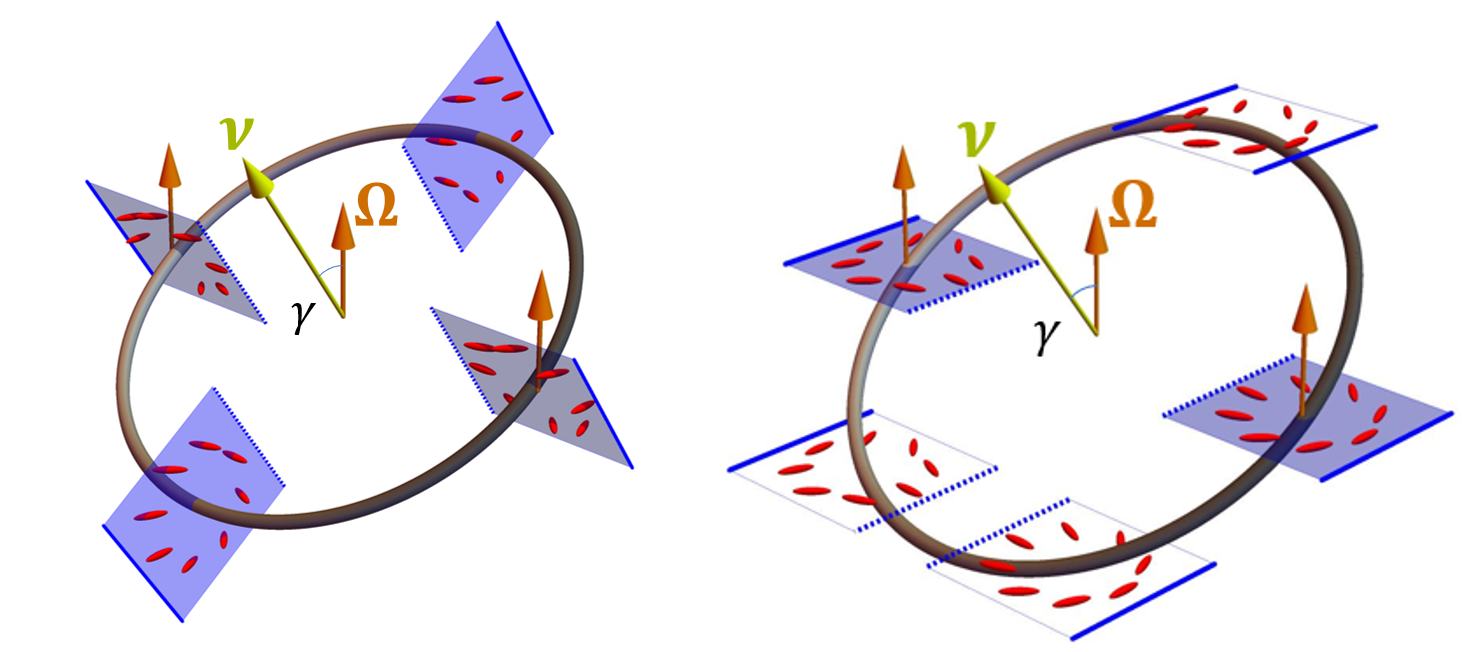}
    \caption{ A schematic illustration of initial disclination loop in the ideal case, visualized by \textbf{left}: normal cross-sections, and \textbf{right}: principal planes.  }\label{fig:compare}
\end{figure}

As shown in the main text,  $\gamma$ averaged over all initial loops in our simulations collapses onto $\ke$, suggesting that wedge-twist loops are formed through the bend instability at low $\ke$, while pure-twist loops (or loops approaching the pure-twist type with $\gamma\sim30^\circ$) emerge at high $\ke$. Although the existing model predicts a significant transformation in loop type with varying $\ke$, our analysis of principal planes indicates that these loops are formed by similar bend-induced mechanisms. The only difference is that, at high $\ke$, the initial loops are additionally influenced by twist, leading to a tilted structure and consequently a reduced $\gamma$. Therefore, while the framework based on normal cross-sections provides a local topological analysis, the principal plane analysis reveals the geometrical structure at a larger length scale through an intuitive quasi-2D perspective, enabling the investigation of the mechanics underlying loop formation. A schematic illustration of an initial disclination loop in the ideal case, comparing the visualizations based on normal cross-sections and principal planes, is shown in Fig.~\ref{fig:compare}. 

\section{Analysis of steady state properties}
\label{sec:appendix_steady}

 This section introduces the algorithm used to calculate the statistical properties of the steady states in our simulation. These properties include $\langle S \rangle$, the correlation function of the order parameter, and the Frank deformation. We also discuss how we determined the master curve on $\kappa/(1+2\fa^2)$.

 \textbf{Mean $S$}: $\langle S \rangle$ is  $S$ spatially averaged throughout the entire system. 

 \textbf{The spatial correlation function}: The spatial correlation function is evaluated using the Wiener–Khinchin theorem under the assumption of translational symmetry in the steady state. Specifically, it is given by:
\begin{equation}
    C_S(\vect{r}) \coloneqq \langle \Delta S(\vect{r'}) \Delta S(\vect{r'}+\vect{r})\rangle_{\vect{r'}}=\frac{1}{\sqrt{2\pi}}\int |\widetilde{\Delta S}(\vect{k})|^2\exp(i\vect{k}\cdot\vect{r})\text{d}\vect{k}
\end{equation}
where $\Delta S = S - \langle S \rangle$.

 To account for the nematic symmetry, we quantify the correlation for directors using $[\nd(\vect{r'})\cdot\nd(\vect{r'}+\vect{r})]^2$ instead of $\nd(\vect{r'})\cdot\nd(\vect{r'}+\vect{r})$. This leads to ($\qn\coloneqq\norm{n}\norm{n}$):
\begin{equation}
    C_{\nd} (\vect{r})\coloneqq\langle\qn(\vect{r'}):\qn(\vect{r}'+\vect{r})|\rangle_{\vect{r}'}=\frac{1}{\sqrt{2\pi}}\int[ \tilde{\qn}(\vect{k}):\tilde{\qn}^\dagger(\vect{k})]\exp(i\vect{k}\cdot\vect{r})\text{d}\vect{k}
\end{equation}
where $\dagger$ denotes complex conjugation.
Assuming rotational symmetry in the steady state, since the effects of initial conditions have relaxed, $C(\vect{r})$ collapses onto a single curve $C(r)$. This serves as one of the criteria for determining whether the system has reached  steady state. Fitting $C(r)$ with exponential decay provides the corresponding correlation length.

 \textbf{Frank deformation}: \textbf{Frank deformation}: Throughout this work, we refer to the expressions involving splay, twist, and bend as the \emph{Frank deformation}, rather than the \emph{Frank energy}. This distinction is intentional: calculating the true Frank free energy requires knowledge of the elastic moduli associated with each deformation mode, which are macroscopic material parameters that we have not calculated for our system. Instead, we evaluate the deformation fields themselves, without applying modulus weighting. 
 
 There are two problems if the expression of Frank deformation is directly used in our simulation configurations where defects exist. First, the traditional expression is defined using $\nd$, which is effectively a vector. This is not valid in nematic systems with defects, as $\nd$ transitions to $-\nd$ when traversing a loop around a defect. Second, the Frank deformation is only valid in ordered regions; in disordered regions around defects, the elastic deformation is influenced by the biaxial structure and  gradients of $S$, and currently there is no way to calculate the Frank deformation in a way that has a clear physical meaning. Moreover, $\nd$ becomes ill-defined as $S$ approaches zero.

To address the first challenge, we employ $\tens{q}$ instead of $\nd$ to compute the Frank deformation. This approach differs slightly from the method in Ref.~\cite{Frank}. First, we use $\tens{q}\coloneqq\norm{n}\norm{n}-\tens{I}/3$ instead of $\tens{Q}$ to exclude the gradient of $S$ and biaxial contributions. Second, rather than directly calculating the Frank deformation, we derive the following quantities and measure their norms:
\begin{align}
    [(\nabla\cdot\nd)\nd]_i  &=  \tens{q}_{ab}\p{a}\tens{q}_{ib} + 2\,\tens{q}_{ia}\p{b}\tens{q}_{ab} \\
    [\nd\times(\nabla\times\nd)]_i &= -2\,\tens{q}_{ab}\p{a}\tens{q}_{ib} - \tens{q}_{ia}\p{b}\tens{q}_{ab} \\
    \nd\cdot(\nabla\times\nd) &= \tens{\epsilon}_{abc}\tens{q}_{ad}\p{b}\tens{q}_{cd}
\end{align}
The first two expressions represent vector quantities, known as the splay and bend vectors, respectively, while  the third expression is a pseudoscalar, referred to as the twist\cite{Selinger_2018}. By analyzing our simulation results, we have found that our method more accurately calculates the Frank deformation compared to the approach in Ref.~\cite{Frank}.

We note that the term \emph{twist} is not always used consistently across the literature. In this work, we define twist as the pseudoscalar quantity characterizing the handedness of the director field, which may be positive or negative. In contrast, the Frank free energy depends on the square of this quantity, which is also often referred to as twist in other contexts. To avoid ambiguity, we emphasize that our use of \emph{twist} here refers to the geometric measure rather than the energy contribution.

 To address the second challenge in disordered regions, we weight the Frank deformation by $S^2$, reducing its significance of regions near defects. This procedure explains the re-entrant behavior observed in the collapse of Frank deformation shown in main text: While the system exhibits small deformations in the high $\kappa$ limit, the low $\kappa$ limit also reduces deformation due to the uniformly low $S$ and the prevalence of defects.

\textbf{Identifying the master curve}: Based on the stress-balance analysis discussed in main text and our previous findings in Ref.~\cite{joshi_interplay_2019}, the calculated quantities, as functions of stiffness and activity, we anticipated that quantities may collapse onto a single curve defined by $\kappa/(1+g \fa^h)$. Here, $g$ and $h$ are empirical parameters, with $h$ theoretically predicted to be approximately 2. Our data analysis indicates that when both $g$ and $h$ are set to around 2, the collapse behavior is generally well-aligned. Since no widely accepted metric exists to quantify the quality of data collapse, we instead determined the values of $g$ and $h$ through an analysis of time scales. As discussed above,  the time scales of the \emph{Instability} stage are well-defined by the onset time of nonlinear dynamics and the time at which the first disclination loop emerges. These two time scales exhibit a linear correlation with $\kappa/(1+g \fa^h)/\fa$. By fitting these time scales with a linear function, we found that $g = h = 2$ were the best-fit parameter values. 


\section{The colormap for 3D directors}
\label{sec:colormap}

 This section introduces the colormap used for the filaments in Fig.~1 in the main text. We have developed and used this colormap for two reasons. First, unlike in 2D nematics, where the orientation is readily perceived, human eyes struggle to recognize the orientation of 3D nematics on paper due to the projection of the 3D directors onto a 2D plane. Thus, a well-designed colormap could significantly help readers understand the 3D structure. Second, an ideal colormap should be a continuous and one-to-one map from the state space of directors into the color space. In 2D, this colormap has been identified and well applied. However,  such an ideal colormap does not exist for 3D directors. Therefore,  designing a sufficiently practical and effective colormap remains a non-trivial task. 

We propose that an optimal colormap should satisfy the following two properties: (1) each unique direction corresponds to a distinct color, and (2) similar directions map to similar colors. The first criterion ensures the uniqueness of directional information, allowing for clear differentiation of orientations within the director field, particularly in regions with complex or rapidly varying orientations. This unique mapping minimizes potential information loss, enabling viewers to intuitively and accurately interpret the field's directional structure. The second criterion enhances continuity in the visualization by capturing the smooth or gradual changes in orientation that typically characterize nematic director fields. By mapping similar directions to proximate colors, the colormap reinforces the perception of directional gradients, enhancing the viewer's ability to discern subtle shifts in orientation.

 A more mathematically rigorous description is as follows. The colormap is defined as a mapping from the space of directors to the space of colors.  The space of orientations of 3D directors is $\mathbb{RP}^2$, while the space of colors, for the case of the RGB color mode, is the unit cuboid domain within 3D Euclidean space, $\mathbb{R}^3$. The two requirements mentioned above can be restated as a need for this map to be bijective and continuous, which implies that we are seeking a homeomorphism from $\mathbb{RP}^2$ to a subspace of $\mathbb{R}^3$. Mathematically, this corresponds to finding a map that \textbf{topologically embeds} $\mathbb{RP}^2$ into $\mathbb{R}^3$.

 However, it has been proven that such an embedding does not exist. An intuitive explanation for why $\mathbb{RP}^2$ can be embedded in $\mathbb{R}^4$ instead of $\mathbb{R}^3$ is as follows. It is well known that biaxial nematics can be continuously described by a symmetric, traceless $3 \times 3$ tensor $\tens{Q}$, implying that the space of biaxial nematics is homeomorphic to a subspace of $\mathbb{R}^5$. Representing a single 3D director via its second moment effectively constrains $\tens{Q}$ to a uniaxial form, thereby reducing one degree of freedom. 

Therefore, we seek to slightly relax the first requirement and keep the second one because the continuity allows viewers to identify gradient structures or defects within the material. Assigning distinct colors to similar directions may introduce visual noise, which could obscure the inherent physical properties of the field.

 Since this work primarily focuses on disclinations, the color representation of directors should exhibit rapid variation around disclination points to reflect the local structure accurately. This requires that any arbitrary path connecting two antipodal points on a unit sphere spans a broad color spectrum, even though some points may map to the same color, as the requirement for one-to-one correspondence is relaxed. Importantly, no such path should map to a single color. \textbf{Immersion} can facilitate this goal by ensuring that each point along the path is locally distinct in color, as immersion is precisely a local embedding. Fortunately, an immersion from $\mathbb{RP}^2$ into $\mathbb{R}^3$ is mathematically feasible. A common example is Boy's surface, where a director represented by $(x,y,z)$ transforms into $(x',y',z')$ as
\setlength{\jot}{10pt}
\begin{align}
x' & = \frac{1}{2} \left[\left(2 x^2-y^2-z^2\right)+x y \left(y^2-x^2\right)+x z \left(x^2-z^2\right)+2 y z \
\left(y^2-z^2\right)\right] \notag \\
y' & =  \frac{7 }{8} \left[\left(y^2-z^2\right) +x y \left(y^2-x^2\right)+x z \left(z^2-x^2\right)\right] \label{eqn:Boy} \\
z' & = \frac{1}{8} (x+y+z) \left[(x+y+z)^3+4 (y-x) (x-z) (z-y)\right] \notag
\end{align}
where $x$, $y$ and $z$ represent the orientation of $\nd$ in the vector representation. We adjust these expressions for the following two goals. 1) Enclose the state space of $(x',y',z')$ into the unit cuboid domain within $\mathbb{R}^3$. 2) Make the color distinct on the white background. 3) Set the color for $x, y$ and $z$ direction closed with red, blue and green, respectively. The final expression is given by $(R, B, G)$ as follows:
\begin{align*}
    x'' &= (61x' - 18y' - 29z')/60 \\
    y'' &= (-61x' - 90y' - 79z')/60 \\
    z'' &= (-11x' + 18y' + 79z')/60 \\
    R &= x''/2.1 + 0.45\\
    B &= y''/4.2 + 0.51\\
    G &= z''/2 + 0.23
\end{align*} 


\section{Supplementary Video Descriptions}
\label{sec:videos}

\begin{itemize}

\item \textbf{Supplementary Video 1 (a-c)}: Animation of a simulation trajectory illustrating the system's dynamics within the entire simulation box, evolving from an initially ordered state to a chaotic steady state. The parameters used are $\fa=1500$, $\kappa=2$, and $\ke=167.7$. The visualized content in each panel is as follows: \textbf{(a)} Active filaments, with colors assigned based on their closest alignment to the Cartesian axes ($x$, $y$, or $z$), following the colormap introduced in section~\ref{sec:videos}. \textbf{(b)} Directors at the boundaries of the simulation box, colored using the same colormap. \textbf{(c)} Disclination loops, with each disclination point represented as a black dot.

\item \textbf{Supplementary Video 2}: Dynamics of disclinations in the steady state, shown in a zoomed  in region of the simulation box with directors on the boundary lines color-coded by their orientation. The directors surrounding the point disclinations are highlighted. Parameters are $\fa=250$, $\kappa=2$, $\ke=27.8$. 
 
\item \textbf{Supplementary Video 3}: Example of scanning a disclination loop with principal planes. Parameters are $\fa=100$, $\kappa=1$, $\ke=33.3$. Directors are projected on the principal planes to form streamlines.


\item \textbf{Supplementary Video 4}: Illustrative animation showing the geometrical structure of initial loops. It highlights how increasing $\ke$ shears the bend structure, inducing additional twist and thereby reducing the angle $\gamma$ between the loop's normal $\norm{\upnu}$ and the rotation vector $\norm{\Omega}$.

\end{itemize}

The relevant data and key functions to produce animations are backed up on~\cite{OSF}.


\section{Backup of data and code}
\label{sec:DataAndCode}
This project is backed up in the GitHub repository~\cite{GitHub}, which includes:
\begin{itemize}
    \item \emph{active-code}: The code that applies tangential active forces on filaments was implemented within the LAMMPS framework. Our simulations were conducted using the LAMMPS version downloaded from the official website on December 26, 2023, with the active force module incorporated into the directory \emph{lammps/src/BROWNIAN}.
    \item \emph{simulation}: The simulation and analysis code performed on cluster. The general package for investigation of 3D nematics is \emph{Nematics3D}.
    \item \emph{Loop-Analysis}: The analysis results of initial loops as the distribution of several variables. 
\end{itemize}

\bibliography{Loops}